\documentclass[journal]{IEEEtran}
 \usepackage{setspace}
 \usepackage{amsfonts,amssymb}
\usepackage{csquotes}
\usepackage{algorithm}

\usepackage{algcompatible}

\newcommand{\newalgname}[1]{%
  \renewcommand{\ALG@name}{#1}%
}

\newcommand{\argmax}{\operatornamewithlimits{argmax}}

\newcommand{\argmin}{\operatornamewithlimits{argmin}}

\usepackage{tikz}

 \usepackage[top=.75in,bottom=.75in,left=.625in,right=.625in]{geometry}
\usepackage{pifont}
\newcommand{\var}[1]{\mbox{$var [#1]$}}
\usepackage{color}
\makeatletter
\newcommand{\vast}{\bBigg@{4}}
\newcommand{\Vast}{\bBigg@{5}}
\makeatother
\usepackage{multirow}
\usepackage{amsmath}
\usepackage{url}
\usepackage{graphicx}
\usepackage{ntheorem}
 \usepackage{mathrsfs}
\usepackage{cite}
\usepackage{enumitem}
 \usepackage{cases}
\newif\ifdraft
\draftfalse 
\newcommand{\fsquare}{\vrule height6pt width7pt depth1pt}
           \newcommand{\myproof}{{\hfill \\ \bf Proof. \ }}           
\newcommand{\myendpf}{\hfill\fsquare \\[0.0in]}
\newcommand{\pfe}{\hfill\fsquare \\[0.0in]}
\ifCLASSINFOpdf
\else
\fi

 \newfont{\secfnta}{ptmb8t at 12pt}

 \usepackage{relsize}

\newcommand{\1}[1]{{\bf 1}\left[#1\right]}

\newcommand{\bcap} {\hspace{2pt} \mathlarger{\cap}
\hspace{2pt}}

\newcommand{\ccap} {\hspace{2pt} \cap
\hspace{2pt}}

    \makeatletter
    \def\@listi{\leftmargin\leftmargini
        \parsep 1\p@ \@plus0\p@ \@minus\p@
        \topsep 2\p@   \@plus0\p@ \@minus\p@
        \itemsep1\p@ \@plus0\p@ \@minus\p@}
    \let\@listI\@listi\@listi
    \makeatother

\newcommand*\mcapinn[2]{\vcenter{\hbox{$\mathsurround=0pt
\ifx\displaystyle#1\textstyle\else#1\fi\bigcap$}}}

\newcommand*\mcupinn[2]{\vcenter{\hbox{$
\bigcup$}}}

\newcommand{\bP}[1]{{\mathbb{P}}\left[{#1}\right]}
\newcommand{\bE}[1]{{\mathbb{E}}\left[{#1}\right]}

\newtheorem{fact}{Fact}
\newtheorem{lem}{Lemma}
\newtheorem{thm}{Theorem}

\newtheorem{rem}{Remark}
\newtheorem{cor}{Corollary}
\newtheorem{proposition}{Proposition}

\begin{document}

\title{Analyzing Connectivity of Heterogeneous\\Secure Sensor Networks}

\author{Jun~Zhao,~\IEEEmembership{Member,~IEEE}\thanks{The author was with the Cybersecurity Lab (CyLab) at Carnegie Mellon University, Pittsburgh, PA 15213, USA. He is now with Arizona State University, Tempe, AZ 85281, USA (Email: junzhao@alumni.cmu.edu).}}
\maketitle

 \pagestyle{empty}
 \thispagestyle{empty}
%


%




\begin{abstract}

We analyze connectivity of a heterogeneous secure sensor network that uses   key predistribution to protect communications between sensors. For this network on a set $\mathcal{V}_n$ of $n$ sensors, suppose there is a pool $\mathcal{P}_n$ consisting of $P_n$ distinct keys. The $n$ sensors in $\mathcal{V}_n$ are divided into $m$ groups $\mathcal{A}_1, \mathcal{A}_2, \ldots, \mathcal{A}_m$. Each sensor $v$ is independently assigned to exactly a group according to the probability distribution with $\mathbb{P}[v \in \mathcal{A}_i]= a_i$ for  $i=1,2,\ldots,m$, where $\sum_{i=1}^{m}a_i  = 1$. 
  Afterwards, each sensor in group $\mathcal{A}_i$ independently chooses $K_{i,n}$ keys uniformly at random from the key pool $\mathcal{P}_n$, where $K_{1,n} \leq K_{2,n}\leq \ldots \leq K_{m,n}$. Finally, any two sensors in $\mathcal{V}_n$ establish a secure link in between if and only if they have at least one key in common. We present critical conditions for connectivity of this heterogeneous secure sensor network. The result provides useful
guidelines for the design of secure sensor networks.

This paper improves the seminal work \cite{yagan-heter-IT} (IEEE Transactions on Information Theory 2016) of Ya{\u{g}}an on connectivity in the following aspects.
First, our result is more broadly applicable; specifically, we consider $K_{m,n} / K_{1,n} = o(\sqrt{n}\hspace{1pt})$, while \cite{yagan-heter-IT} requires $K_{m,n} / K_{1,n} = o(\ln n)$. Put differently,
$K_{m,n} / K_{1,n}$ in our paper examines the case of $\Theta(n^{x})$ for any $x <\frac{1}{2}$ and $\Theta\big((\ln n)^y\big)$ for any $y>0$, while that of \cite{yagan-heter-IT} does not cover any $\Theta(n^{x})$, and covers $\Theta\big((\ln n)^y)$ for only $0<y<1$.  This improvement is possible due to a delicate coupling argument. Second, although both studies show that a  critical scaling for connectivity is that the \vspace{1pt} term $b_n$ denoting $\sum_{j=1}^{m} \left\{ a_j \left[1 -{{{P_n-K_{1,n}}\choose{K_{j,n}}}\big/{{P_n}\choose{K_{j,n}}}}\right] \right\}$ equals \vspace{1pt} $\frac{\ln n}{n}$, our paper considers any of $b_{n}=o\big(\frac{ \ln  n }{n}\big)$, $b_{n}=\Theta\big(\frac{ \ln  n }{n}\big)$, and  \vspace{.5pt} $b_{n}=\omega\big(\frac{ \ln  n }{n}\big)$, while \cite{yagan-heter-IT} evaluates  \vspace{1pt} only $b_{n}=\Theta\big(\frac{ \ln  n }{n}\big)$. Third, in terms of characterizing the transitional behavior of  \vspace{1.5pt} connectivity, our scaling $b_{n}=\frac{ \ln  n + \beta_n}{n}$ for a sequence $\beta_n$ is more fine-grained than the scaling $b_{n}\sim \frac{c \ln  n}{n}$ for a constant $c \neq 1$ of \cite{yagan-heter-IT}. In a nutshell, we add the case of $c=1$ in $b_{n}\sim \frac{c \ln  n}{n}$,  \vspace{.5pt} where the graph can be
 connected or disconnected asymptotically, depending on the limit
of $\beta_n$.

Finally, although a recent study by Eletreby and Ya{\u{g}}an \cite{yagan-heter-k} uses the fine-grained scaling discussed above for a more complex graph model, their result (just like \cite{yagan-heter-IT}) also demands $K_{m,n} / K_{1,n} = o(\ln n)$, which is less general than $K_{m,n} / K_{1,n} = o(\sqrt{n}\hspace{1pt})$ addressed in this paper.

\end{abstract}
%

%
%
%

%

\begin{IEEEkeywords}

Secure sensor networks, heterogeneity, connectivity, key predistribution.
  \end{IEEEkeywords}

\section{Introduction}

\subsection{Modeling secure sensor networks}

\textbf{Securing wireless sensor networks via key predistribution.}
Wireless sensor networks (WSNs) enable a broad range
of applications including military surveillance, patient monitoring, and home automation \cite{zhaoconnectivityTAC,iyengar2016distributed,virgil}. In many
cases, WSNs are deployed in hostile environments (e.g., battlefields),
making it crucial to use cryptographic protection to
secure sensor communications. To that end, significant efforts
have been devoted to developing strategies for securing WSNs,
and random key predistribution schemes have been broadly
accepted as promising solutions.

The idea of  key predistribution initiated by Eschenauer and Gligor \cite{virgil} is that cryptographic keys are assigned to sensors before deployment to ensure secure communications after deployment. The Eschenauer--Gligor (EG) scheme
\cite{virgil} works as follows. For a  WSN of $n$
sensors, in the key predistribution phase, a large
\emph{key pool} $\mathcal{P}_n$ consisting of  $P_n$ different cryptographic keys is used to
select \emph{uniformly at random} $K_n$ distinct keys for each
sensor node. These $K_n$ keys constitute the \emph{key ring} of a
sensor, and are installed in the sensor's memory. After deployment,
two sensors establish secure communication over a wireless link if
and only if their key rings have at least one key in common. Common
keys are found in the neighbor discovery phase whereby a random
constant is enciphered in all keys of a node and broadcast along
with the resulting ciphertext block. The key pool size ${P}_n$ and the key ring size
$K_n$ are both functions of $n$ in order to consider the scaling behavior. The
condition $1 \leq K_n \leq P_n$ holds naturally.

\textbf{Random key graphs.} A secure sensor network under the EG  scheme described above induces the so-called \textit{random key graph} $\mathbb{G}(n,K_n,P_n)$ \cite{ISIT,ZhaoYaganGligor,yagan,virgillncs}. In this graph of $n$ nodes, each node selects $K_n$ keys uniformly at random from a common key pool $\mathcal{P}_n$ of $P_n$ keys, and two nodes establish an undirected edge in between if and only if they share at least one key. Random key graphs (also known as uniform  random intersection graph \cite{r1,r10,bloznelis2013,zhaoconnectivityTAC}) have received significant interest recently with applications beyond secure WSNs; e.g., recommendation systems \cite{r4},
clustering and classification \cite{GodehardtJaworski,bloznelis2013,GodehardtJaworskiRybarczyk}, cryptanalysis of hash functions \cite{r10}, frequency hopping \cite{ZhaoAllerton}, and
the modeling of epidemics \cite{ball2014}.

\subsection{Modeling heterogeneous secure sensor networks} \label{section-intro-model}

\textbf{Heterogeneous secure sensor networks.} The EG  scheme above assigns the same number of keys to each sensor. Yet, in practice, sensors may have varying levels of memory and computational
resources. In view of this heterogeneity, we study a variation \cite{yagan-heter-IT} of the EG
scheme that is more suitable for \emph{heterogeneous secure sensor networks} \cite{du2007effective,lu2008framework,hussain2007efficient}. In this scheme \cite{yagan-heter-IT},  the key ring size of each sensor is independently drawn from  $\overrightarrow{K}:=[K_{1,n},\ldots, K_{m,n}]$
according to a probability vector $\overrightarrow{a}:=[a_1,\ldots,a_m]$ (i.e., $K_{i,n}$ is taken with probability $a_{i}$ for  $i=1,2,\ldots,m$), where $m$ is a positive constant integer, and $a_i|_{i=1,2,\ldots,m}$ are positive constants satisfying the natural condition $\sum_{i=1}^{m}a_i  = 1$ (note that $m$ and $a_i|_{i=1,2,\ldots,m}$ do not scale with $n$). The above process can also be understood as follows: for  $i=1,2,\ldots,m$, each sensor first joins a group $\mathcal{A}_i$ with probability $a_i$; after assigning to a particular group $\mathcal{A}_i$, a sensor independently chooses $K_{i,n}$ different keys uniformly at random from a common pool $\mathcal {P}_n$
of $P_n$ distinct  keys.


\textbf{Heterogeneous random key graphs.} \vspace{1pt}
We let $\mathbb{G}(n,\overrightarrow{a}_n,\overrightarrow{K_n},P_n)$ denote the
  graph topology of a heterogeneous secure sensor network employing the above key predistribution
scheme, and refer to   this graph as a \emph{heterogeneous random key graph}. Formally, it is defined on  a set $\mathcal{V}_n$ of $n$ nodes as follows. All nodes are divided into $m$ different groups $\mathcal{A}_1, \mathcal{A}_2, \ldots, \mathcal{A}_m$. Each node $v\in \mathcal{V}_n$ is independently assigned to exactly one group according to the following probability distribution\footnote{{We summarize the notation and convention} as follows. Throughout the paper, $\bP{\cdot}$ denotes a probability and $\bE{\cdot}$ stands for the expectation of a random variable. All limiting statements are understood with $n \to \infty$. We use the standard
asymptotic notation $o(\cdot), O(\cdot), \omega(\cdot),\Omega(\cdot),
\Theta(\cdot), \sim$; see \cite[Page 2-Footnote 1]{ZhaoYaganGligor} for their meanings. In
particular, ``$\sim$'' represents {asymptotic equivalence} and is  defined as follows: for two positive sequences $f_n$ and $g_n$, the relation $f_n \sim
g_n$ means $\lim_{n \to
  \infty} ( {f_n}/{g_n})=1$. Also, ``$\ln $'' stands for the natural logarithm function, and ``$|\cdot|$'' can denote the absolute value as well as the
cardinality of a set. \label{footnote1}}: $\bP{v \in \mathcal{A}_i}= a_i$ for $i=1,2,\ldots,m$. The edge set is built as follows. To begin with, assume that there exists a pool $\mathcal{P}_n$ consisting of $P_n$ distinct keys. Then for $i=1,2,\ldots,m$, each node in group $\mathcal{A}_i$ independently chooses $K_{i,n}$ different keys uniformly at random from the key pool $\mathcal{P}_n$, where $1\leq K_{i,n}\leq P_n$. Finally, any two nodes in $\mathcal{V}_n$ have an undirected edge in between if and only if they share at least one key.

\subsection{Results and Discussions} \label{sec-ResultsDiscussions}

 For a heterogeneous random key graph $\mathbb{G}(n,\overrightarrow{a}, \overrightarrow{K_n},P_n)$ modeling a heterogeneous secure sensor network,  we establish Theorem \ref{thm:OneLaw+NodeIsolation} below, which improves the pioneering result of Ya\u{g}an \cite{yagan-heter-IT}.

\begin{thm}  \label{thm:OneLaw+NodeIsolation}

Consider a heterogeneous random key
graph $\mathbb{G}(n,\overrightarrow{a}, \overrightarrow{K_n},P_n)$ under $ P_n =
\Omega(n)$ and
\begin{align}
\omega(\sqrt{P_n/n}) = K_{1,n} \leq K_{2,n}\leq \ldots \leq K_{m,n} = o(\sqrt{P_n}).   \label{KPPremo}
\end{align}
With a sequence $\beta_n$ for all $n$ defined by
\begin{align}
\sum_{j=1}^{m} \left\{ a_j \left[1 -{{{P_n-K_{1,n}}\choose{K_{j,n}}}\over{{P_n}\choose{K_{j,n}}}}\right] \right\} & = \frac{ \ln  n + \beta_n}{n},   \label{eq:scalinglaw_old_2}
\end{align}
it holds that
\begin{subnumcases}
{ \hspace{-15pt}  \lim_{n \rightarrow \infty }\hspace{-1pt} \mathbb{P}\hspace{-1pt}\bigg[
\hspace{-3pt}\begin{array}{c}
\mathbb{G}(n,\overrightarrow{a}, \overrightarrow{K_n},P_n) \\
\mbox{is connected.}
\end{array}\hspace{-3pt}
\bigg] \hspace{-2pt}=\hspace{-2pt}}  \hspace{-3pt}0,\quad\hspace{-4pt}\text{if  }\lim_{n \to \infty}{\beta_n} =- \infty, \label{thm-con-eq-0c} \\
\hspace{-3pt}1,\quad\hspace{-4pt}\text{if  }\lim_{n \to \infty}{\beta_n} = \infty. \label{thm-con-eq-1c}
\end{subnumcases}
\end{thm}

\textbf{A sharp zero--one law of connectivity.} Theorem \ref{thm:OneLaw+NodeIsolation} presents a \textit{sharp} zero--one law, since the zero-law (\ref{thm-con-eq-0c}) shows that the
graph is  connected \emph{almost surely} under certain parameter conditions while the one-law (\ref{thm-con-eq-1c}) shows that the
graph is disconnected  \emph{almost surely} if parameters are slightly changed,
where an event (indexed by $n$) {occurs \emph{almost surely} if its probability converges to $1$ as
$n\to \infty$}.

\textbf{Improvements over Ya\u{g}an \cite{yagan-heter-IT}.} This paper improves the seminal work \cite{yagan-heter-IT} of Ya{\u{g}}an on connectivity in the following aspects.


\begin{itemize}
\item[\textbf{(i)}] \textbf{\textit{More practical conditions.}} Our result is more broadly applicable; specifically, from (\ref{KPPremo}), we consider $K_{m,n} / K_{1,n} = o(\sqrt{n}\hspace{1pt})$, while \cite{yagan-heter-IT} requires $K_{m,n} / K_{1,n} = o(\ln n)$. Put differently,
$K_{m,n} / K_{1,n}$ in our paper examines the case of $\Theta(n^{x})$ for any $x <\frac{1}{2}$ and $\Theta\big((\ln n)^y\big)$ for any $y>0$, while that of \cite{yagan-heter-IT} does not cover any $\Theta(n^{x})$, and covers $\Theta\big((\ln n)^y)$ for only $0<y<1$.  This improvement is possible due to a delicate coupling argument. See Algorithm \ref{alg-find-K} on Page \pageref{alg-find-K} as an illustration for the difficulty of the argument.
\item[\textbf{(ii)}]  \textbf{\textit{More fine-grained zero--one law.}} Both this paper and \cite{yagan-heter-IT} show that a  critical scaling for connectivity is that the term $b_n$ denoting the left hand side of (\ref{eq:scalinglaw_old_2}) equals $\frac{\ln n}{n}$. However, in terms of characterizing the transitional behavior \vspace{1pt} of connectivity, our scaling $b_{n}=\frac{ \ln  n + \beta_n}{n}$ for a sequence $\beta_n$ \vspace{1pt} is more fine-grained than the scaling $b_{n}\sim \frac{c \ln  n}{n}$ for a constant $c \neq 1$ of \cite{yagan-heter-IT}. In a nutshell, we add the case of $c=1$ in $b_{n}\sim \frac{c \ln  n}{n}$, where the graph can be
 connected or disconnected asymptotically, depending on the limit
of $\beta_n$.
\item[\textbf{(iii)}]  \textbf{\textit{More general scaling condition.}} Our paper considers  any of $b_{n}=o\big(\frac{ \ln  n }{n}\big)$, $b_{n}=\Theta\big(\frac{ \ln  n }{n}\big)$, and $b_{n}=\omega\big(\frac{ \ln  n }{n}\big)$, \vspace{2pt} while \cite{yagan-heter-IT} evaluates  only $b_{n}=\Theta\big(\frac{ \ln  n }{n}\big)$
\end{itemize}

\textbf{Improvements over Eletreby and Ya{\u{g}}an \cite{yagan-heter-k1,yagan-heter-k}.} Although a recent research by Eletreby and Ya{\u{g}}an \cite{yagan-heter-k} uses the fine-grained scaling discussed above for a more complex graph model (another work \cite{yagan-heter-k1} by them uses the weaker scaling), both studies \cite{yagan-heter-k1,yagan-heter-k} (just like \cite{yagan-heter-IT}) also demand $K_{m,n} / K_{1,n} = o(\ln n)$, which is less general than $K_{m,n} / K_{1,n} = o(\sqrt{n}\hspace{1pt})$ addressed in this paper.

\textbf{Improvements over Zhao \emph{et al.} \cite{zhaoconnectivityTAC}.}  Recently, Zhao \emph{et al.} \cite{zhaoconnectivityTAC}   consider $k$-connectivity of heterogeneous random key graphs, where $k$-connectivity means that connectivity is still preserved despite the deletion of at most $(k-1)$ arbitrary nodes. Although  $k$-connectivity of \cite{zhaoconnectivityTAC} is stronger than our connectivity, their result applies to only a narrow range of parameters since it only permits a very small  variance of the
key ring sizes.



\textbf{Interpreting (\ref{eq:scalinglaw_old_2}).}
From \cite{yagan-heter-IT} as well as the  explanation later in Section \ref{sec:Preliminaries}, the left hand side of the scaling condition (\ref{eq:scalinglaw_old_2}) is in fact the  {mean}
probability of edge occurrence for a group-$1$ node (i.e., a node in group $\mathcal{A}_{1}$), where the mean is taken by considering that the other endpoint of the edge can fall into each group $\mathcal{A}_{j}$ with probability $a_{j}$ for  $j=1,2,\ldots,m$.

\subsection{Organization}

We organize the remainder of the paper as follows. Section \ref{related} presents related work. Then we introduce experiments in Section \ref{sec:expe} to confirm our theoretical result (i.e., Theorem \ref{thm:OneLaw+NodeIsolation}).  Afterwards, we provide proof details for Theorem \ref{thm:OneLaw+NodeIsolation} in Sections \ref{sec:Preliminaries}--\ref{sec-prf-thm-under-cpl}. Finally,  we conclude the paper in Section \ref{sec:Conclusion}.


\section{Related Work} \label{related}

 {Random key graphs} have received significant interest recently with applications spanning secure sensor networks  \cite{r1,Rybarczyk,DiPietroTissec,adrian,virgil}, recommender systems \cite{r4},
clustering and classification \cite{GodehardtJaworski,bloznelis2013}, cryptanalysis \cite{r10}, and
 epidemics \cite{ball2014}. Random key graphs are also referred to as  {uniform~random~intersection~graphs} in the literature   \cite{r4,bloznelis2013,r10,zhaoconnectivityTAC}, where the word ``uniform'' is due to the fact that in a random key graph $\mathbb{G}(n,K_n,P_n)$, the number of keys assigned to  each node is fixed as $K_n$ given $n$. The graph $\mathbb{G}(n,K_n,P_n)$ has been   studied in terms of connectivity \cite{DiPietroTissec,r1,ryb3,yagan}, $k$-connectivity \cite{ISIT,ZhaoYaganGligor}, $k$-robustness \cite{ZhaoCDC,zhaoconnectivityTAC}, component evolution \cite{Rybarczyk}, clustering coefficient \cite{bloznelis2013}, and diameter \cite{ryb3}.

\begin{figure}[!t]
 \hspace{-10pt} \includegraphics[width=.5\textwidth]{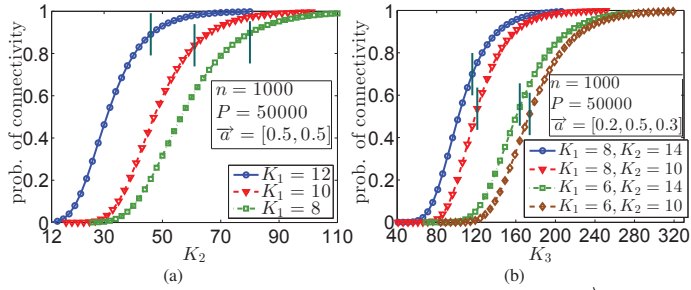}
\vspace{-20pt}\caption{We plot the
connectivity probability of
graph $\mathbb{G}(n,\protect\overrightarrow{a}, \protect\overrightarrow{K},P)$ when $\protect\overrightarrow{K}=[K_1,K_2]$ varies in Figure \ref{fig1-2}-(a), and when $\protect\overrightarrow{K}=[K_1,K_2,K_3]$ varies in Figure \ref{fig1-2}-(b). In Figure \ref{fig1-2}-(a) (resp., \ref{fig1-2}-(b)), each vertical line presents the
minimal $K_2$ (resp., $K_3$) such that $b_1(\protect\overrightarrow{a},\protect\overrightarrow{K},P)$ in Eq. (\ref{b1noscaling}) is at least $\frac{\ln n}{n}$.\vspace{0pt}} \label{fig1-2}
\end{figure}

\begin{figure}[!t]
  \hspace{-22pt}\includegraphics[width=.55\textwidth]{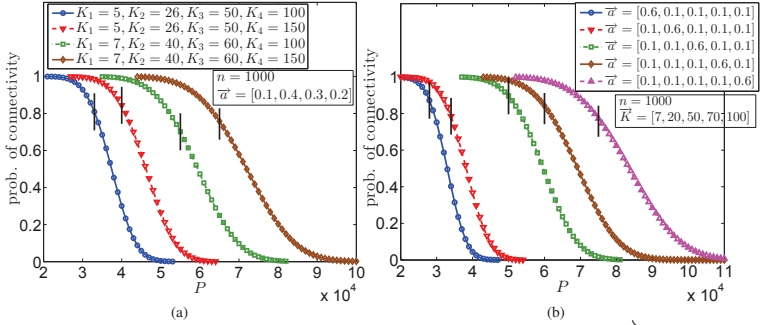} \vspace{-20pt}\caption{We plot the
connectivity probability of
graph $\mathbb{G}(n,\protect\overrightarrow{a}, \protect\overrightarrow{K},P)$ when $P$ varies given different $\protect\overrightarrow{K}$ in Figure \ref{fig3-4}-(a), and given different $\protect\overrightarrow{a}$ in \ref{fig3-4}-(b). Each vertical line presents the
maximal $P$ such that $b_1(\protect\overrightarrow{a},\protect\overrightarrow{K},P)$ in Eq. (\ref{b1noscaling}) is at least $\frac{\ln n}{n}$.\vspace{0pt}} \label{fig3-4}
\end{figure}

\begin{figure}[!t]
   \hspace{-10pt}\includegraphics[width=.52\textwidth]{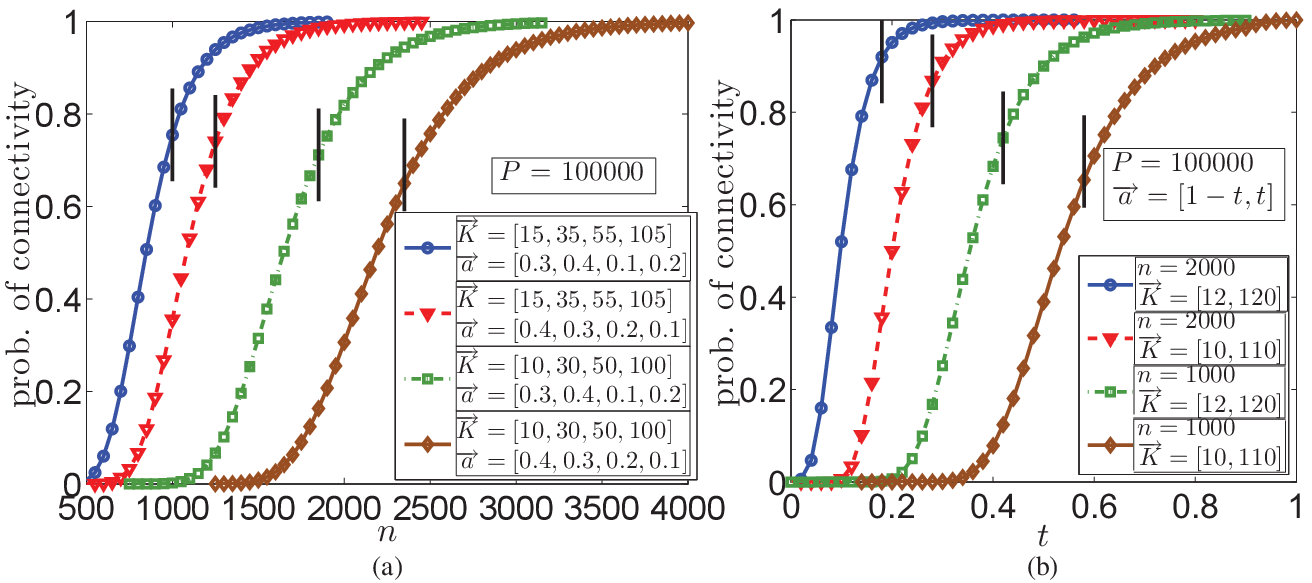}  \vspace{-20pt}\caption{We plot the
connectivity probability of
graph $\mathbb{G}(n,\protect\overrightarrow{a}, \protect\overrightarrow{K},P)$ when $n$ varies in Figure \ref{fig3-4}-(a), and when $\protect\overrightarrow{a}=[1-t, t]$ varies in Figure \ref{fig3-4}-(b). In Figure \ref{fig5-6}-(a) (resp., \ref{fig5-6}-(b)), each vertical line presents the
minimal $n$ (resp., $t$) such that $b_1(\protect\overrightarrow{a},\protect\overrightarrow{K},P)$ in Eq. (\ref{b1noscaling}) is at least $\frac{\ln n}{n}$.\vspace{10pt}} \label{fig5-6}
\end{figure}

In this paper, we study the heterogeneous random key graph model $\mathbb{G}(n,\overrightarrow{a}, \overrightarrow{K_n},P_n)$ \cite{yagan-heter-IT}, where nodes can have different numbers of keys. This graph models a heterogeneous sensor network where sensors have varying level of resources. Compared with the seminal work \cite{yagan-heter-IT} of Ya{\u{g}}an (and its conference version \cite{yagan2016connectivity-ISIT}) on connectivity,
our work has the following improvements, as already discussed in Section \ref{sec-ResultsDiscussions} above (we do not repeat the details here): more practical conditions by considering $K_{m,n} / K_{1,n} = o(\sqrt{n}\hspace{1pt})$ instead of just $K_{m,n} / K_{1,n} = o(\ln n)$, more fine-grained zero--one law by considering the scaling $ \frac{ \ln  n + \beta_n}{n}$ rather than $ \frac{c \ln  n}{n}$ for \vspace{1pt} a constant $c \neq 1$, and more general scaling condition. Although a recent work by Eletreby and Ya{\u{g}}an \cite{yagan-heter-k} uses the fine-grained scaling discussed above for a more complex graph model (other researches \cite{yagan-heter-k1,osyCDC} by them uses the weaker scaling), all these studies \cite{yagan-heter-k1,yagan-heter-k,osyCDC} (just like \cite{yagan-heter-IT}) still demand $K_{m,n} / K_{1,n} = o(\ln n)$, which is less general than $K_{m,n} / K_{1,n} = o(\sqrt{n}\hspace{1pt})$ addressed in this paper. In addition,  Goderhardt \emph{et al.} \cite{GodehardtJaworski,GodehardtJaworskiRybarczyk} and Zhao \emph{et al.} \cite{zhaoconnectivityTAC,ZhaoCDC} also study heterogeneous random key graphs, but their results apply to only a narrow range of parameters and are not applicable to practical secure sensor networks. Finally,
   Bloznelis \emph{et al.} \cite{Rybarczyk} investigate  component evolution rather than connectivity and present   conditions for the existence of a giant connected component (i.e., a connected component of $\Theta(n)$ nodes).

 For heterogeneous secure sensor networks, different key management schemes \cite{du2007effective,hussain2007efficient,lu2008framework} have been proposed, but existing connectivity analyses for them are   informal. In this paper, we formally analyze connectivity and improve \cite{yagan-heter-IT} for heterogeneous secure sensor networks under a simple variant of the Eschenauer--Gligor key predistribution scheme \cite{virgil}.

\section{Experimental Results}
\label{sec:expe}

We now present experimental results   to confirm our theoretical findings of connectivity in
graph $\mathbb{G}(n,\protect\overrightarrow{a}, \protect\overrightarrow{K},P)$, where we write $\protect\overrightarrow{K_n},P_n$ as $\protect\overrightarrow{K},P$ to suppress the subscript $n$.

In Figure \ref{fig1-2}-(a), we plot the
connectivity probability of
graph $\mathbb{G}(n,\protect\overrightarrow{a}, \protect\overrightarrow{K},P)$ for $\protect\overrightarrow{K}=[K_1,K_2]$, with respect to $K_2$ given different $K_1$ (all parameters are provided in the figure). In Figure \ref{fig1-2}-(a) as well as all other figures, for each data point, we
generate $1000$ independent samples of $\mathbb{G}(n,\protect\overrightarrow{a}, \protect\overrightarrow{K},P)$, record the count that the obtained graph is
connected, and then divide the count by $1000$ to obtain the corresponding empirical probability of network connectivity. From the plot, we   see the evident transitional behavior of connectivity. Furthermore, in Figure \ref{fig1-2}-(a) as well as all other figures, based on (\ref{eq:scalinglaw_old_2}) of Theorem \ref{thm:OneLaw+NodeIsolation}, we illustrate the parameter such that $\frac{\ln n}{n}$ roughly equals the left hand side of (\ref{eq:scalinglaw_old_2}), which we denote by $b_1(\protect\overrightarrow{a},\protect\overrightarrow{K},P)$ after suppressing the subscript $n$ (we use $b_1$ in consistence with later notation); i.e.,
\begin{align}
b_1(\protect\overrightarrow{a},\protect\overrightarrow{K},P) : = \sum_{j=1}^{m} \left\{ a_j \cdot \left[1- {{{P-K_{1}}\choose{K_{j}}}\over{{P}\choose{K_{j}}}}\right]\right\}. \label{b1noscaling}
\end{align}
Specifically, in Figure \ref{fig1-2}-(a), each vertical line presents the
minimal $K_2$ such that $b_1(\protect\overrightarrow{a},\protect\overrightarrow{K},P)$ in  (\ref{b1noscaling}) with $\protect\overrightarrow{K}=[K_1,K_2]$ is at least $\frac{\ln n}{n}$.

In Figure \ref{fig1-2}-(b), we plot the
connectivity probability of
graph $\mathbb{G}(n,\protect\overrightarrow{a}, \protect\overrightarrow{K},P)$ for $\protect\overrightarrow{K}=[K_1,K_2,K_3]$ with respect to $K_3$ given different $K_1,K_2$, and each vertical line presents the
minimal $K_3$ such that $b_1(\protect\overrightarrow{a},\protect\overrightarrow{K},P)$ in  (\ref{b1noscaling}) is at least $\frac{\ln n}{n}$.

In Figure \ref{fig3-4}-(a) (resp., \ref{fig3-4}-(b)), we plot the
connectivity probability of
graph $\mathbb{G}(n,\protect\overrightarrow{a}, \protect\overrightarrow{K},P)$ for $\protect\overrightarrow{K}=[K_1,K_2,K_3,K_4]$ with respect to $P$ given different $\protect\overrightarrow{K}$ (resp., given different $\protect\overrightarrow{a}$), and each vertical line presents the
maximal $P$ such that $b_1(\protect\overrightarrow{a},\protect\overrightarrow{K},P)$ in  (\ref{b1noscaling}) is at least $\frac{\ln n}{n}$.

In Figure \ref{fig5-6}-(a), we plot the
connectivity probability of
graph $\mathbb{G}(n,\protect\overrightarrow{a}, \protect\overrightarrow{K},P)$ with respect to $n$ given different $\protect\overrightarrow{K}$ and $\protect\overrightarrow{a}$, and each vertical line presents the
minimal $n$ such that $b_1(\protect\overrightarrow{a},\protect\overrightarrow{K},P)$ in   (\ref{b1noscaling}) is at least $\frac{\ln n}{n}$. In Figure \ref{fig5-6}-(b), we plot the
connectivity probability of
graph $\mathbb{G}(n,\protect\overrightarrow{a}, \protect\overrightarrow{K},P)$ for $\protect\overrightarrow{a}=[1-t, t]$ with respect to $t$ given different $n$ and $\protect\overrightarrow{K}$, and each vertical line presents the
minimal $t$ such that $b_1(\protect\overrightarrow{a},\protect\overrightarrow{K},P)$ in  (\ref{b1noscaling}) is at least $\frac{\ln n}{n}$.

In all figures, we clearly see the transitional behavior of connectivity, and that the transition happens when $b_1(\protect\overrightarrow{a},\protect\overrightarrow{K},P)$ in  (\ref{b1noscaling}) is around $\frac{\ln n}{n}$.
  Summarizing the above, the experiments have confirmed our analytical results.

 \section{Preliminaries}
\label{sec:Preliminaries}

We notate the $n$ nodes in graph $\mathbb{G}(n,\overrightarrow{a}, \overrightarrow{K_n},P_n)$ by $v_1,
v_2, \ldots, v_n$; i.e., $\mathcal{V}_n = \{v_1,
v_2, \ldots, v_n \}$.  For each
$x=1,2\ldots,n$, the  set of keys on node $v_x$ is denoted by $S_x$. When $v_x$ belongs to a group $\mathcal{A}_i$ for some $i \in \{1,2,\ldots,m\}$, the set $S_x$ is uniformly distributed among all
$K_{i,n}$-size subsets of the object pool $\mathcal{P}_n$.

In graph $\mathbb{G}(n,\overrightarrow{a}, \overrightarrow{K_n},P_n)$, let $E_{xy}$ be the event that two different nodes $v_x$ and $v_y$ have an edge in between. Clearly, $E_{xy}$ is equivalent to the event $S_{x} \cap S_{y} \neq \emptyset$. To analyze $E_{xy}$, we often condition on the case where $v_x$ belongs to group $\mathcal{A}_i$ and $v_y$ belongs to group $\mathcal{A}_j$, where $i \in \{1,2,\ldots,m\}$ and $j \in \{1,2,\ldots,m\}$ (note that $x$ and $y$ are different, but $i$ and $j$ may be the same; i.e., different nodes $v_x$ and $v_y$ may belong to the same group).

We define {$p_{i,j,n}$ as the
probability of edge occurrence between a group-$i$ node and a group-$j$ node}. More formally, $p_{i,j,n}$ equals the probability that an edge exists between nodes $v_x$ and $v_y$ conditioning on that  $v_x$ belongs to group $\mathcal{A}_i$ and $v_y$ belongs to group $\mathcal{A}_j$. We now compute $p_{i,j,n}   : =  \mathbb{P} [E_{xy}~|~(v_x\in\mathcal{A}_i)\ccap (v_y\in\mathcal{A}_j) ].$ Let $T(K_{i,n},P_n)$ be the set of all $K_{i,n}$-size subsets of the object pool $\mathcal{P}_n$. Under $(v_x\in\mathcal{A}_i)\ccap (v_y\in\mathcal{A}_j)$, the set $S_x$ (resp., $S_y$) is uniformly distributed in $T(K_{i,n},P_n)$ (resp., $T(K_{j,n},P_n)$). Let $S_x^*$ be an arbitrary element in $T(K_{i,n},P_n)$. Conditioning on $S_x=S_x^{\mathlarger{*}}$, the event $\overline{E_{xy}}$ (i.e., $S_{x} \cap S_{y} = \emptyset$) means $S_{y}\subseteq \mathcal{P}_n \setminus S_x^{\mathlarger{*}}$. Noting that there are ${P_n}\choose{K_{j,n}}$ ways to select a $K_{j,n}$-size set from $\mathcal{P}_n$ and there are ${P_n-K_{i,n}}\choose{K_{j,n}}$ ways to select a $K_{j,n}$-size set from $\mathcal{P}_n\setminus S_x^{\mathlarger{*}}$, we   obtain $\bP{\overline{E_{xy}}~|~(S_x=S_x^{\mathlarger{*}})\ccap (v_y\in\mathcal{A}_j)}  = {{{P_n-K_{i,n}}\choose{K_{j,n}}}\big/{{P_n}\choose{K_{j,n}}}}$.
Given the above, we derive
 \begin{align}
p_{i,j,n} & =  \sum_{S_x^{\mathlarger{*}} \in T(K_{i,n},P_n)} \nonumber \\ &   \Big\{\hspace{-1pt} \bP{S_x\hspace{-1pt}=\hspace{-1pt}S_x^{\mathlarger{*}}\,|\,v_x\hspace{-1pt}\in\hspace{-1pt} \mathcal{A}_i}  \bP{{E_{xy}}\,|\,(S_x\hspace{-1pt}=\hspace{-1pt}S_x^{\mathlarger{*}})\cap (v_y\hspace{-1pt}\in\hspace{-1pt}\mathcal{A}_j)} \hspace{-2pt}\Big\} \nonumber \\ & = 1- {{{P_n-K_{i,n}}\choose{K_{j,n}}}\over{{P_n}\choose{K_{j,n}}}}, \label{psq1contaxadrst}
\end{align}
where we use $\sum_{S_x^{\mathlarger{*}} \in T(K_{i,n},P_n)} \bP{S_x=S_x^{\mathlarger{*}}~|~ v_x\in \mathcal{A}_i}=1$.


%
%
%

We further define {$b_{i,n}$ as  the {mean}
probability of edge occurrence for a group-$i$ node}. More formally, $b_{i,n}$ is the probability that an edge exists between nodes $v_x$ and $v_y$ conditioning on that  $v_x$ belongs to group $\mathcal{A}_i$. Since $v_y$ belongs to group $\mathcal{A}_{j}$ with probability $a_{j}$ for $j=1,\ldots,n$, we have $b_{i,n}  = \sum_{j=1}^{m} \big( a_j p_{i,j,n}\big)$. From this and (\ref{psq1contaxadrst}), we can see that $b_{1,n}$ (i.e., $b_{i,n}$ with $i=1$) equals the left hand side of (\ref{eq:scalinglaw_old_2}) in our Theorem \ref{thm:OneLaw+NodeIsolation}; namely,
 \begin{align}
b_{1,n} & = \sum_{j=1}^{m} \left\{ a_j \left[1 -{{{P_n-K_{1,n}}\choose{K_{j,n}}}\over{{P_n}\choose{K_{j,n}}}}\right] \right\} .\label{psq1conta7tttaareae-restatedeqn} 
\end{align}
Although the above results are also discussed in \cite{yagan-heter-IT}, we present them clearly here for better understanding.

\section{Confining $|\beta_n|$ as $o(\ln n)$ in Theorem \ref{thm:OneLaw+NodeIsolation}} \label{sec-confine-beta-n-thm:OneLaw+NodeIsolation-old}

We recall from (\ref{eq:scalinglaw_old_2}) that $\beta_n$ measures the deviation of the left hand side of (\ref{eq:scalinglaw_old_2}) from the critical scaling $\frac{ \ln n }{n}$. The desired results (\ref{thm-con-eq-0c}) and (\ref{thm-con-eq-1c}) of Theorem \ref{thm:OneLaw+NodeIsolation} consider $\lim_{n \to \infty}{\beta_n} =- \infty$ and $\lim_{n \to \infty}{\beta_n} = \infty$, respectively. In principle, the absolute value $|\beta_n|$ can be arbitrary as long as it is unbounded. Yet, we
 will explain that
the extra condition $|{\beta_n} |=  o ( \ln n)$
can be introduced in proving
Theorem \ref{thm:OneLaw+NodeIsolation}. Specifically, we will show
\begin{align}
\hspace{-17pt}\begin{array}{l}\text{Theorem \ref{thm:OneLaw+NodeIsolation} with the additional condition $|\beta_n|= o(\ln n)$}   \\  \Longrightarrow ~~
\text{Theorem \ref{thm:OneLaw+NodeIsolation} regardless of $|\beta_n|= o(\ln n)$}.\end{array}
\label{with_extra}
\end{align}

We write $b_{1,n}$ in (\ref{psq1conta7tttaareae-restatedeqn}) as $b_{1}(\overrightarrow{a}, \overrightarrow{K_n},P_n)$. Given
$\overrightarrow{a}, \overrightarrow{K_n},P_n$, one can determine
$\beta_n$ from (\ref{eq:scalinglaw_old_2}).
 In order to show (\ref{with_extra}), we   present Lemma \ref{lem_Gq_cplinga} below.


\begin{lem} \label{lem_Gq_cplinga}
 {For a graph $\mathbb{G}(n,\overrightarrow{a}, \overrightarrow{K_n},P_n)$ on a probability space $\mathbb{S}$ under
\begin{align}
 P_n =
\Omega(n),\label{PnOmeganeval}
\end{align}
and
\begin{align}
\omega(\sqrt{P_n/n}) = K_{1,n} \leq K_{2,n}\leq \ldots \leq K_{m,n} = o(\sqrt{P_n}),\label{K1n-K2n-Kmn-condition}
\end{align}
with a sequence $\beta_n$ defined by
\begin{align}
b_1(\overrightarrow{a}, \overrightarrow{K_n},P_n)=\frac{ \ln  n + \beta_n}{n},\label{lem-coupling-condition-edge-prob-beta-n}
\end{align}
the following results hold:
\begin{itemize}[leftmargin=10pt]
\item[(i)]
If
\begin{align}
{\lim_{n \to \infty}\beta_n=-\infty}\,,\label{condition-beta-n-minus-infinity}
\end{align}
there exists a graph $\mathbb{G}(n,\overrightarrow{a}, \overrightarrow{K_n^{\mathlarger{*}}},P_n)$ on the probability space $\mathbb{S}$ such that $\mathbb{G}(n,\overrightarrow{a}, \overrightarrow{K_n},P_n)$ is a spanning \textbf{subgraph} of $\mathbb{G}(n,\overrightarrow{a}, \overrightarrow{K_n^{\mathlarger{*}}},P_n)$, where   $$\omega(\sqrt{P_n/n}) = K_{1,n}^{\mathlarger{*}} \leq K_{2,n}^{\mathlarger{*}}\leq \ldots \leq K_{m,n}^{\mathlarger{*}} = o(\sqrt{P_n})$$ and a sequence $\beta_n^{\mathlarger{*}}$ defined by
\begin{align}
b_1(\overrightarrow{a}, \overrightarrow{K_n^{\mathlarger{*}}},P_n)=\frac{ \ln  n + \beta_n^{\mathlarger{*}}}{n} \label{lem-coupling-condition-edge-prob-beta-n-star}
\end{align}
 satisfies ${\lim_{n \to \infty}{\beta_n^{\mathlarger{*}}} =- \infty\text{ and }|\beta_n^{\mathlarger{*}}|= o(\ln n)}$\,.
\item[(ii)] If
\begin{align}
{\lim_{n \to \infty}\beta_n=\infty}\,,\label{condition-beta-n-infinity}
\end{align}
there exists a graph $\mathbb{G}(n,\overrightarrow{a}, \overrightarrow{K_n^{\#}},P_n)$ on the probability space $\mathbb{S}$ such that $\mathbb{G}(n,\overrightarrow{a}, \overrightarrow{K_n},P_n)$ is a spanning \textbf{supergraph} of $\mathbb{G}(n,\overrightarrow{a}, \overrightarrow{K_n^{\#}},P_n)$, where   $$\omega(\sqrt{P_n/n}) = K_{1,n}^{\#} \leq K_{2,n}^{\#}\leq \ldots \leq K_{m,n}^{\#} = o(\sqrt{P_n})$$ and a sequence $\beta_n^{\#}$ defined by\begin{align}
b_1(\overrightarrow{a}, \overrightarrow{K_n^{\#}},P_n)=\frac{ \ln  n + \beta_n^{\#}}{n} \label{lem-coupling-condition-edge-prob-beta-n-pound}
\end{align}
 satisfies ${\lim_{n \to \infty}{\beta_n^{\#}} =  \infty\text{ and }|\beta_n^{\#}|= o(\ln n)}$\,.
\end{itemize}
 }

\end{lem}

Before establishing Lemma \ref{lem_Gq_cplinga}, we first demonstrate (\ref{with_extra})  given Lemma \ref{lem_Gq_cplinga}.

\subsubsection{\textbf{Proving (\ref{with_extra})  given Lemma \ref{lem_Gq_cplinga}}}~


To establish (\ref{with_extra}) using Lemma \ref{lem_Gq_cplinga}, we discuss the two cases in the result of Theorem \ref{thm:OneLaw+NodeIsolation} below:
\ding{172} $\lim_{n \to \infty}\beta_n = -\infty$, and \ding{173} $\lim_{n \to \infty}\beta_n = \infty$.

\ding{172} Under $\lim_{n \to \infty}\beta_n = -\infty$, we use the property (i) of Lemma \ref{lem_Gq_cplinga} to have graph $\mathbb{G}(n,\overrightarrow{a}, \overrightarrow{K_n^{\mathlarger{*}}},P_n)$. Then if
Theorem \ref{thm:OneLaw+NodeIsolation} holds with the additional condition $|\beta_n|= o(\ln n)$, we apply the zero-law (\ref{thm-con-eq-0c}) of Theorem \ref{thm:OneLaw+NodeIsolation} to graph $\mathbb{G}(n,\overrightarrow{a}, \overrightarrow{K_n^{\mathlarger{*}}},P_n)$ and obtain that this graph is disconnected almost surely, which implies that its spanning subgraph $\mathbb{G}(n,\overrightarrow{a}, \overrightarrow{K_n},P_n)$ is also disconnected almost surely. This means that the zero-law (\ref{thm-con-eq-0c}) of Theorem \ref{thm:OneLaw+NodeIsolation} holds   regardless of $|\beta_n|= o(\ln n)$.

\ding{173} Under $\lim_{n \to \infty}\beta_n = \infty$, we use the property (ii) of Lemma \ref{lem_Gq_cplinga} to have graph $\mathbb{G}(n,\overrightarrow{a}, \overrightarrow{K_n^{\#}},P_n)$. Then if
Theorem \ref{thm:OneLaw+NodeIsolation} holds with the additional condition $|\beta_n|= o(\ln n)$, we apply the one-law (\ref{thm-con-eq-1c}) of Theorem \ref{thm:OneLaw+NodeIsolation} to graph $\mathbb{G}(n,\overrightarrow{a}, \overrightarrow{K_n^{\#}},P_n)$ and obtain that this graph is connected almost surely, which implies that its spanning supergraph $\mathbb{G}(n,\overrightarrow{a}, \overrightarrow{K_n},P_n)$ is also connected almost surely. This means that the one-law (\ref{thm-con-eq-1c}) of Theorem \ref{thm:OneLaw+NodeIsolation} holds   regardless of $|\beta_n|= o(\ln n)$.

\subsubsection{\textbf{Proving Lemma \ref{lem_Gq_cplinga}}}~

 \textbf{Proving Property (i) of Lemma \ref{lem_Gq_cplinga}:} \vspace{2pt}

 We
 define $\widetilde{\beta_n^{\mathlarger{*}}}$ by
 \begin{align}
\widetilde{\beta_n^{\mathlarger{*}}} &  = \max\{\beta_n, -\ln \ln n\}.\label{al2-parta}
\end{align}
Since $1 -{{{P_n-K_{1,n}}\choose{X}}\over{{P_n}\choose{X}}}$ is the probability that a node with key ring size $X$ and a node with key ring size $K_{1,n}$ have an edge in between when their key rings are independent selected uniformly at random from the same pool of $P_n$ keys, it is increasing as $X$ increases. This can also be formally shown through ${{{P_n-K_{1,n}}\choose{X+1}}\over{{P_n}\choose{X+1}}} \Bigg/\Bigg[{{{P_n-K_{1,n}}\choose{X}}\over{{P_n}\choose{X}}}\Bigg]= 1 - \frac{K_{1,n}}{P_n-X} < 1$. Then we
define ${K}_{m,n}^{\mathlarger{*}}$ as the \textit{maximal} non-negative integer $X$ such that
\begin{align}
a_m \left[1 -{{{P_n-K_{1,n}}\choose{X}}\over{{P_n}\choose{X}}}\right] + \sum_{j=1}^{m-1} \left\{ a_j \left[1 -{{{P_n-K_{1,n}}\choose{K_{j,n}}}\over{{P_n}\choose{K_{j,n}}}}\right] \right\}, \label{func-object-1}
\end{align}
is no greater than
\begin{align}
\frac{\ln  n + \widetilde{\beta_n^{\mathlarger{*}}}}{n}; \label{func-object-1-bound}
\end{align}
i.e.,
\begin{align}
{K}_{m,n}^{\mathlarger{*}}:= \argmax\{X: \text{(\ref{func-object-1})} \leq \text{(\ref{func-object-1-bound})}\}. \label{Kmn-star-def}
\end{align}
Such ${K}_{m,n}^{\mathlarger{*}}$  always exists because setting $X$ as ${K}_{m,n}$ induces $\text{(\ref{func-object-1})} \leq \text{(\ref{func-object-1-bound})}$, which follows from (\ref{psq1conta7tttaareae-restatedeqn}) and (\ref{al2-parta}).

We will prove Property (i) of Lemma \ref{lem_Gq_cplinga} by using ${K}_{m,n}^{\mathlarger{*}}$ above and setting ${K}_{j,n}^{\mathlarger{*}}$ as ${K}_{j,n}$ for $1\leq j \leq m-1$; i.e.,
\begin{align}
{K}_{j,n}^{\mathlarger{*}}: ={K}_{j,n},\text{ for $1\leq j \leq m-1$}. \label{set-K-j-n-star}
\end{align}
To this end, we will show the following results:
\begin{itemize}
\item[\textbf{(i.1)}] $\mathbb{G}(n,\overrightarrow{a}, \overrightarrow{K_n},P_n)$ is a spanning subgraph of $\mathbb{G}(n,\overrightarrow{a}, \overrightarrow{K_n^{\mathlarger{*}}},P_n)$.
\item[\textbf{(i.2)}] $K_{1,n}^{\mathlarger{*}} = \omega(\sqrt{P_n/n})$,
\item[\textbf{(i.3)}] $ K_{1,n}^{\mathlarger{*}} \leq K_{2,n}^{\mathlarger{*}}\leq \ldots \leq K_{m,n}^{\mathlarger{*}}$,
\item[\textbf{(i.4)}]  ${K}_{m,n}^{\mathlarger{*}}$ defined by (\ref{Kmn-star-def}) satisfies $K_{m,n}^{\mathlarger{*}} = o(\sqrt{P_n})$, \vspace{1pt}
\item[\textbf{(i.5)}]  $\beta_n^{\mathlarger{*}}$ defined by (\ref{lem-coupling-condition-edge-prob-beta-n-star}) (i.e., $b_1(\overrightarrow{a}, \overrightarrow{K_n^{\mathlarger{*}}},P_n)=\frac{ \ln  n + \beta_n^{\mathlarger{*}}}{n}$)
  satisfies $\lim_{n \to \infty}{\beta_n^{\mathlarger{*}}} =- \infty$ and $|\beta_n^{\mathlarger{*}}|= o(\ln n)$.
\end{itemize}

We now establish the above results (i.1)--(i.5). 

\textit{\textbf{Proving result  (i.1):}}
We note from (\ref{set-K-j-n-star}) that ${K}_{j,n}^{\mathlarger{*}} ={K}_{j,n}$ for $1\leq j \leq m-1$, and note from (\ref{Kmn-Kmn-star}) that ${K}_{m,n} \leq {K}_{m,n}^{\mathlarger{*}}$. Then from the construction of $\mathbb{G}(n,\overrightarrow{a}, \overrightarrow{K_n},P_n)$ and $\mathbb{G}(n,\overrightarrow{a}, \overrightarrow{K_n^{\mathlarger{*}}},P_n)$, result  (i.1) clearly follows.

\textit{\textbf{Proving results (i.2) and (i.3):}}

Since (\ref{psq1conta7tttaareae-restatedeqn}) and (\ref{al2-parta}) together imply that setting $X$ as ${K}_{m,n}$ induces $\text{(\ref{func-object-1})} \leq \text{(\ref{func-object-1-bound})}$, we obtain from (\ref{Kmn-star-def}) that
\begin{align}
{K}_{m,n} \leq {K}_{m,n}^{\mathlarger{*}}. \label{Kmn-Kmn-star}
\end{align}
Combining (\ref{set-K-j-n-star}) (\ref{Kmn-Kmn-star}) and the condition (\ref{K1n-K2n-Kmn-condition}) (which enforces $\omega(\sqrt{P_n/n}) = K_{1,n} \leq K_{2,n}\leq \ldots \leq K_{m,n} $), we clearly obtain $\omega(\sqrt{P_n/n}) = K_{1,n}^{\mathlarger{*}} \leq K_{2,n}^{\mathlarger{*}}\leq \ldots \leq K_{m,n}^{\mathlarger{*}}$; i.e., results (i.2) and (i.3) are proved.

\textit{\textbf{Proving result  (i.4):}}

Applying the condition (\ref{condition-beta-n-minus-infinity}) (i.e., $\lim_{n \to \infty}\beta_n=-\infty$) and $\lim_{n \to \infty}(-\ln \ln n)=-\infty$ to (\ref{al2-parta}), we obtain
\begin{align}
 \lim_{n \to \infty}\widetilde{\beta_n^{\mathlarger{*}}} =- \infty . \label{widetilde-al2-partalimit}
\end{align}
From $\lim_{n \to \infty}\beta_n=-\infty$,
it holds that $\beta_n \leq 0$ for all $n$ sufficiently large. Then from (\ref{al2-parta}), we have
\begin{align}
 \widetilde{\beta_n^{\mathlarger{*}}} = - O(\ln \ln n) = - o(\ln n),  \label{widetilde-al2-parta}
\end{align}
Setting $X$ as ${K}_{m,n}^{\mathlarger{*}}$ in (\ref{func-object-1}), we use (\ref{func-object-1-bound})  (\ref{Kmn-star-def}) and (\ref{widetilde-al2-parta}) (i.e., $\widetilde{\beta_n^{\mathlarger{*}}}   = -O(\ln \ln n) \leq o(\ln  n)$) to obtain
\begin{align}
&a_m \left[1 -{{{P_n-K_{1,n}}\choose{{K}_{m,n}^{\mathlarger{*}}}}\over{{P_n}\choose{{K}_{m,n}^{\mathlarger{*}}}}}\right] + \sum_{j=1}^{m-1} \left\{ a_j \left[1 -{{{P_n-K_{1,n}}\choose{K_{j,n}}}\over{{P_n}\choose{K_{j,n}}}}\right] \right\} \label{b1aKnPn-expr} \\  & \leq \frac{\ln  n + \widetilde{\beta_n^{\mathlarger{*}}}}{n} \leq \frac{\ln  n}{n} \times [1+o(1)] ,\label{aph5-parta}
\end{align}
which further implies
\begin{align}
 a_m \left[1 -{{{P_n-K_{1,n}}\choose{{K}_{m,n}^{\mathlarger{*}}}}\over{{P_n}\choose{{K}_{m,n}^{\mathlarger{*}}}}}\right]  & = O\bigg(\frac{\ln n}{n}\bigg) .\label{aph5-parta-further}
\end{align}

Since $a_m$ is a positive constant, (\ref{aph5-parta-further})  induces
\begin{align}
1 -{{{P_n-K_{1,n}}\choose{{K}_{m,n}^{\mathlarger{*}}}}\over{{P_n}\choose{{K}_{m,n}^{\mathlarger{*}}}}} & = O\bigg(\frac{\ln n}{n}\bigg) .\label{aph5-parta-Kmbound}
\end{align}

The left hand side of (\ref{aph5-parta-Kmbound}) is the probability that a node with key ring size ${K}_{m,n}^{\mathlarger{*}}$ and a node with key ring size $K_{1,n}$ have an edge in between when their key rings are independent selected uniformly at random from the same pool of $P_n$ keys. Then (\ref{aph5-parta-Kmbound}) and \cite[Lemma 4.2]{yagan-heter-IT} together imply
\begin{align}
\frac{K_{1,n} {K}_{m,n}^{\mathlarger{*}}}{P_n} & \sim 1 -{{{P_n-K_{1,n}}\choose{{K}_{m,n}^{\mathlarger{*}}}}\over{{P_n}\choose{{K}_{m,n}^{\mathlarger{*}}}}} ,\label{aph5-parta-Kmbound-equiv-sb}
\end{align}
which along with (\ref{aph5-parta-Kmbound})   gives
\begin{align}
\frac{K_{1,n} {K}_{m,n}^{\mathlarger{*}}}{P_n} & = O\bigg(\frac{\ln n}{n}\bigg).\label{aph5-parta-Kmbound-equiv}
\end{align}
Then (\ref{aph5-parta-Kmbound-equiv}) and  $K_{1,n} = \omega(\sqrt{P_n/n})$ (from (\ref{K1n-K2n-Kmn-condition})) further induces
\begin{align}
{K}_{m,n}^{\mathlarger{*}} & = O\bigg(\frac{\ln n}{n}\bigg) \cdot \frac{P_n}{K_{1,n}}\nonumber \\ & = O\bigg(\frac{\ln n}{n}\bigg) \cdot o\bigg(\frac{P_n}{\sqrt{P_n/n}}\bigg)= \sqrt{P_n} \cdot \frac{\ln n}{\sqrt{n}} = o(\sqrt{P_n});\nonumber
\end{align}
i.e., result  (i.4) is proved.

\textit{\textbf{Proving result  (i.5):}}

To prove result  (i.5), we will bound $b_1(\overrightarrow{a}, \overrightarrow{K_n^{\mathlarger{*}}},P_n)$. From (\ref{set-K-j-n-star}), the only difference between $\overrightarrow{K_n^{\mathlarger{*}}}$ and $\overrightarrow{K_n}$ is that the $m$th dimension of $\overrightarrow{K_n^{\mathlarger{*}}}$ is ${K}_{m,n}^{\mathlarger{*}}$, while the $m$th dimension of $\overrightarrow{K_n}$ is ${K}_{m,n}$. Then replacing ${K}_{m,n}$ by ${K}_{m,n}^{\mathlarger{*}}$ in the expression of $b_1(\overrightarrow{a}, \overrightarrow{K_n},P_n)$ in (\ref{psq1conta7tttaareae-restatedeqn}), we obtain that
$b_1(\overrightarrow{a}, \overrightarrow{K_n^{\mathlarger{*}}},P_n)$ equals the term in (\ref{b1aKnPn-expr}); i.e.,
\begin{align}
&b_1(\overrightarrow{a}, \overrightarrow{K_n^{\mathlarger{*}}},P_n)\nonumber \\ & = a_m \left[1 -{{{P_n-K_{1,n}}\choose{{K}_{m,n}^{\mathlarger{*}}}}\over{{P_n}\choose{{K}_{m,n}^{\mathlarger{*}}}}}\right] + \sum_{j=1}^{m-1} \left\{ a_j \left[1 -{{{P_n-K_{1,n}}\choose{K_{j,n}}}\over{{P_n}\choose{K_{j,n}}}}\right] \right\} \label{b1aKnPn-expr2}
\end{align}
As proved in (\ref{aph5-parta}), it holds that
\begin{align}
&b_1(\overrightarrow{a}, \overrightarrow{K_n^{\mathlarger{*}}},P_n) \leq \frac{\ln  n + \widetilde{\beta_n^{\mathlarger{*}}}}{n}\leq  \frac{\ln n}{n} \cdot [1+o(1)].\label{b1aKnPn-expr2-dev1}
\end{align}
(\ref{b1aKnPn-expr2-dev1}) gives an upper bound for $b_1(\overrightarrow{a}, \overrightarrow{K_n^{\mathlarger{*}}},P_n)$.
We now further provide a lower bound for $b_1(\overrightarrow{a}, \overrightarrow{K_n^{\mathlarger{*}}},P_n)$. To this end, we  observe that we can first evaluate the probability when we change $\overrightarrow{K_n^{\mathlarger{*}}}$ in $b_1(\overrightarrow{a}, \overrightarrow{K_n^{\mathlarger{*}}},P_n)$ such that the $m$th dimension of $\overrightarrow{K_n^{\mathlarger{*}}} :=[K_{1,n}^{\mathlarger{*}},\ldots, K_{m,n}^{\mathlarger{*}}] $ increases by $1$ (i.e., increases to ${K}_{m,n}^{\mathlarger{*}}+1$). More specifically, with $\overrightarrow{L_n^{\mathlarger{*}}}$ defined by $$\overrightarrow{L_n^{\mathlarger{*}}}:= [K_{1,n}^{\mathlarger{*}},\ldots, K_{m-1,n}^{\mathlarger{*}}, K_{m,n}^{\mathlarger{*}}+1] , $$ we evaluate $b_1(\overrightarrow{a}, \overrightarrow{L_n^{\mathlarger{*}}},P_n)$. From (\ref{set-K-j-n-star}), we further have $\overrightarrow{L_n^{\mathlarger{*}}} = [K_{1,n} ,\ldots, K_{m-1,n} , K_{m,n}^{\mathlarger{*}}+1] $. Then replacing ${K}_{m,n}$ by ${K}_{m,n}^{\mathlarger{*}}+1$ in the expression of $b_1(\overrightarrow{a}, \overrightarrow{K_n},P_n)$ in (\ref{psq1conta7tttaareae-restatedeqn}), we obtain $b_1(\overrightarrow{a}, \overrightarrow{L_n^{\mathlarger{*}}},P_n)$ via
\begin{align}
&b_1(\overrightarrow{a}, \overrightarrow{L_n^{\mathlarger{*}}},P_n) \nonumber \\ & = a_m \left[1 -{{{P_n-K_{1,n}}\choose{{K}_{m,n}^{\mathlarger{*}}+1}}\over{{P_n}\choose{{K}_{m,n}^{\mathlarger{*}}}+1}}\right] + \sum_{j=1}^{m-1} \left\{ a_j \left[1 -{{{P_n-K_{1,n}}\choose{K_{j,n}}}\over{{P_n}\choose{K_{j,n}}}}\right] \right\} .\label{b1aKnPn-expr2-dev2}
\end{align}
Given the above expression (\ref{b1aKnPn-expr2-dev2}) of $b_1(\overrightarrow{a}, \overrightarrow{L_n^{\mathlarger{*}}},P_n)$, we obtain from the definition of ${K}_{m,n}^{\mathlarger{*}}$ in (\ref{Kmn-star-def}) that
\begin{align}
b_1(\overrightarrow{a}, \overrightarrow{L_n^{\mathlarger{*}}},P_n)  & > \frac{\ln  n + \widetilde{\beta_n^{\mathlarger{*}}}}{n} .\label{aph5-parta-theother}
\end{align}
Given (\ref{aph5-parta-theother}), to bound $b_1(\overrightarrow{a}, \overrightarrow{K_n^{\mathlarger{*}}},P_n)$, we evaluate $ b_1(\overrightarrow{a}, \overrightarrow{L_n^{\mathlarger{*}}},P_n) - b_1(\overrightarrow{a}, \overrightarrow{K_n^{\mathlarger{*}}},P_n)$. From (\ref{b1aKnPn-expr2}) and (\ref{b1aKnPn-expr2-dev2}), it follows that
\begin{align}
& b_1(\overrightarrow{a}, \overrightarrow{L_n^{\mathlarger{*}}},P_n) - b_1(\overrightarrow{a}, \overrightarrow{K_n^{\mathlarger{*}}},P_n)  \nonumber \\ & = a_m \left\{\left[1 -{{{P_n-K_{1,n}}\choose{{K}_{m,n}^{\mathlarger{*}}+1}}\over{{P_n}\choose{{K}_{m,n}^{\mathlarger{*}}}+1}}\right]-\left[ 1 -{{{P_n-K_{1,n}}\choose{{K}_{m,n}^{\mathlarger{*}}}}\over{{P_n}\choose{{K}_{m,n}^{\mathlarger{*}}}}} \right]\right\} \label{b1aKnPn-expr2-dev3-interm} .
\end{align}

To further analyze (\ref{b1aKnPn-expr2-dev3-interm}), we now evaluate $1 -{{{P_n-K_{1,n}}\choose{{K}_{m,n}^{\mathlarger{*}}}}\over{{P_n}\choose{{K}_{m,n}^{\mathlarger{*}}}}}$ and $1 -{{{P_n-K_{1,n}}\choose{{K}_{m,n}^{\mathlarger{*}}+1}}\over{{P_n}\choose{{K}_{m,n}^{\mathlarger{*}}}+1}}$, respectively.

First, (\ref{aph5-parta-Kmbound-equiv}) and \cite[Lemma 4.2]{yagan-heter-IT} together imply
\begin{align}
\begin{array}{l} \mathlarger{1 -{{{P_n-K_{1,n}}\choose{{K}_{m,n}^{\mathlarger{*}}}}\over{{P_n}\choose{{K}_{m,n}^{\mathlarger{*}}}}}  = \frac{K_{1,n} {K}_{m,n}^{\mathlarger{*}}}{P_n} \cdot [1 + x_n^{\mathlarger{*}}]} \\[14pt] \text{ for some }x_n^{\mathlarger{*}} = \pm o(1) . \end{array}\label{aph5-parta-Kmbound-equiv-theother-org}
\end{align}
Second, we now analyze $\frac{K_{1,n} ({K}_{m,n}^{\mathlarger{*}}+1)}{P_n}$, which is useful to evaluate $1 -{{{P_n-K_{1,n}}\choose{{K}_{m,n}^{\mathlarger{*}}+1}}\over{{P_n}\choose{{K}_{m,n}^{\mathlarger{*}}}+1}}$, as will become clear soon. To this end, we first use (\ref{aph5-parta-Kmbound-equiv}) and $K_{1,n}\leq {K}_{m,n} \leq {K}_{m,n}^{\mathlarger{*}}$ (which holds from $K_{1,n}\leq {K}_{m,n}$ of (\ref{K1n-K2n-Kmn-condition}),  and ${K}_{m,n} \leq {K}_{m,n}^{\mathlarger{*}}$ of (\ref{Kmn-Kmn-star})) to obtain
\begin{align}
\frac{{K_{1,n}}^2}{P_n} \leq \frac{K_{1,n} {K}_{m,n}^{\mathlarger{*}}}{P_n} & = O\bigg(\frac{\ln n}{n}\bigg),\label{aph5-parta-Kmbound-equiv-K1n}
\end{align}
so that (\ref{aph5-parta-Kmbound-equiv-K1n}) along with $K_{1,n} = \omega(1)$ (which holds from $P_n =
\Omega(n)$ of (\ref{PnOmeganeval}), and $K_{1,n} =
\omega(\sqrt{P_n/n})$ of (\ref{K1n-K2n-Kmn-condition})) further implies
\begin{align}
 \frac{K_{1,n} }{P_n} = \frac{{K_{1,n}}^2}{P_n}\bigg/K_{1,n} \leq  {O\bigg(\frac{\ln n}{n}\bigg)}\bigg/{\omega(1)} & = o\bigg(\frac{\ln n}{n}\bigg).\label{aph5-parta-Kmbound-equiv-K1n2}
\end{align}
From (\ref{aph5-parta-Kmbound-equiv}) and (\ref{aph5-parta-Kmbound-equiv-K1n2}), it follows that
\begin{align}
\frac{K_{1,n} ({K}_{m,n}^{\mathlarger{*}}+1)}{P_n} & = O\bigg(\frac{\ln n}{n}\bigg) .\label{aph5-parta-Kmbound-equiv-theother-equiv}
\end{align}
Then (\ref{aph5-parta-Kmbound-equiv-theother-equiv}) and \cite[Lemma 4.2]{yagan-heter-IT} together imply
\begin{align}
\begin{array}{l} \mathlarger{ 1 -{{{P_n-K_{1,n}}\choose{{K}_{m,n}^{\mathlarger{*}}+1}}\over{{P_n}\choose{{K}_{m,n}^{\mathlarger{*}}}+1}} = \frac{K_{1,n} ({K}_{m,n}^{\mathlarger{*}}+1)}{P_n} \cdot [1 + y_n^{\mathlarger{*}}]} \\[14pt] \text{for some }y_n^{\mathlarger{*}} = \pm o(1) . \end{array}\label{aph5-parta-Kmbound-equiv-theother-equivb}
\end{align}
The combination of (\ref{aph5-parta-Kmbound-equiv-theother-org}) and (\ref{aph5-parta-Kmbound-equiv-theother-equivb}) yields
\begin{align}
 & \left[1 -{{{P_n-K_{1,n}}\choose{{K}_{m,n}^{\mathlarger{*}}+1}}\over{{P_n}\choose{{K}_{m,n}^{\mathlarger{*}}}+1}}\right]-\left[ 1 -{{{P_n-K_{1,n}}\choose{{K}_{m,n}^{\mathlarger{*}}}}\over{{P_n}\choose{{K}_{m,n}^{\mathlarger{*}}}}} \right] \label{aph5-parta-Kmbound-equiv-theother-equivcsb}\\ & = \frac{K_{1,n} ({K}_{m,n}^{\mathlarger{*}}+1)}{P_n} \cdot [1 + y_n^{\mathlarger{*}}] - \frac{K_{1,n} {K}_{m,n}^{\mathlarger{*}}}{P_n} \cdot [1 + x_n^{\mathlarger{*}}] \nonumber \\ & = \frac{K_{1,n} {K}_{m,n}^{\mathlarger{*}}}{P_n} \cdot [y_n^{\mathlarger{*}} - x_n^{\mathlarger{*}}] + \frac{K_{1,n}}{P_n} \cdot [1 + y_n^{\mathlarger{*}}]  .\label{aph5-parta-Kmbound-equiv-theother-equivc}
\end{align}
From  $\frac{K_{1,n} {K}_{m,n}^{\mathlarger{*}}}{P_n}  = O\big(\frac{\ln n}{n}\big) $ in (\ref{aph5-parta-Kmbound-equiv}), $  \frac{K_{1,n} }{P_n}  \leq o\big(\frac{\ln n}{n}\big) $ in (\ref{aph5-parta-Kmbound-equiv-K1n2}), $x_n^{\mathlarger{*}} = \pm o(1)$ in (\ref{aph5-parta-Kmbound-equiv-theother-org}), $y_n^{\mathlarger{*}} = \pm o(1)$ in (\ref{aph5-parta-Kmbound-equiv-theother-equivb}), we obtain that the right hand side of (\ref{aph5-parta-Kmbound-equiv-theother-equivc}) can be written as $ \pm o\big(\frac{\ln n}{n}\big)$. This result along with the obvious fact that (\ref{aph5-parta-Kmbound-equiv-theother-equivcsb}) is non-negative, implies that (\ref{aph5-parta-Kmbound-equiv-theother-equivcsb}) can be written as $ o\big(\frac{\ln n}{n}\big)$. Then using $\text{(\ref{aph5-parta-Kmbound-equiv-theother-equivcsb})} = o\big(\frac{\ln n}{n}\big)$ and $0<a_m \leq 1$ in (\ref{b1aKnPn-expr2-dev3-interm}), we obtain
\begin{align}
 b_1(\overrightarrow{a}, \overrightarrow{L_n^{\mathlarger{*}}},P_n) - b_1(\overrightarrow{a}, \overrightarrow{K_n^{\mathlarger{*}}},P_n)   &  = o\bigg(\frac{\ln n}{n}\bigg) .\label{b1aKnPn-expr2-dev3}
\end{align}
 From (\ref{aph5-parta-theother}) and (\ref{b1aKnPn-expr2-dev3}), it follows that
\begin{align}
b_1(\overrightarrow{a}, \overrightarrow{K_n^{\mathlarger{*}}},P_n) &  = b_1(\overrightarrow{a}, \overrightarrow{L_n^{\mathlarger{*}}},P_n) -  o\bigg(\frac{\ln n}{n}\bigg) \nonumber \\ & > \frac{\ln  n + \widetilde{\beta_n^{\mathlarger{*}}} - o(\ln  n)}{n}.\label{b1aKnPn-expr2-dev4}
\end{align}
Then from (\ref{b1aKnPn-expr2-dev1}) and (\ref{b1aKnPn-expr2-dev4}), $\beta_n^{\mathlarger{*}}$ defined by (\ref{lem-coupling-condition-edge-prob-beta-n-star}) (i.e., $b_1(\overrightarrow{a}, \overrightarrow{K_n^{\mathlarger{*}}},P_n)=\frac{ \ln  n + \beta_n^{\mathlarger{*}}}{n}$)
  satisfies
  \begin{align}
\widetilde{\beta_n^{\mathlarger{*}}} - o(\ln  n) <  {\beta_n}^{\mathlarger{*}} \leq \widetilde{\beta_n^{\mathlarger{*}}} ,\label{b1aKnPn-expr2-dev6}
\end{align}
Finally, we use (\ref{widetilde-al2-partalimit}) and (\ref{b1aKnPn-expr2-dev6}) to derive $\lim_{n \to \infty}{\beta_n^{\mathlarger{*}}} =- \infty$, and use  (\ref{widetilde-al2-parta}) and (\ref{b1aKnPn-expr2-dev6}) to derive $\beta_n^{\mathlarger{*}} = -o(\ln n)$ so that $|\beta_n^{\mathlarger{*}}|= o(\ln n)$. Hence, result (i.5) is proved.

To summarize, we have established the above results  (i.1)--(i.5), respectively. Then Property (i) of Lemma \ref{lem_Gq_cplinga}  follows immediately.

 \textbf{Proving Property (ii)  of Lemma \ref{lem_Gq_cplinga}:} \vspace{2pt}

 We construct $\protect\overrightarrow{K_n^{\#}}:= [{K}_{1,n}^{\#}, {K}_{2,n}^{\#}, \ldots, {K}_{m,n}^{\#}]$ using Algorithm \ref{alg-find-K}. Our goal here is to prove that such vector $\protect\overrightarrow{K_n^{\#}}$ satisfies Property (ii) of Lemma \ref{lem_Gq_cplinga}. More specifically, we will show the following results:
\begin{itemize}
\item[\textbf{(ii.1)}] $\mathbb{G}(n,\overrightarrow{a}, \overrightarrow{K_n},P_n)$ is a spanning supergraph of $\mathbb{G}(n,\overrightarrow{a}, \overrightarrow{K_n^{\#}},P_n)$.
\item[\textbf{(ii.2)}] $ K_{1,n}^{\#} =\omega(\sqrt{P_n/n})$,
\item[\textbf{(ii.3)}] $  K_{1,n}^{\#} \leq K_{2,n}^{\#}\leq \ldots \leq K_{m,n}^{\#}$,
\item[\textbf{(ii.4)}] $K_{m,n}^{\#} = o(\sqrt{P_n})$, \vspace{1pt}
\item[\textbf{(ii.5)}]  $\beta_n^{\#}$ defined by (\ref{lem-coupling-condition-edge-prob-beta-n-pound}) (i.e., $b_1(\overrightarrow{a}, \overrightarrow{K_n^{\#}},P_n)=\frac{ \ln  n + \beta_n^{\#}}{n}$)
  satisfies $\lim_{n \to \infty}{\beta_n^{\#}} =  \infty$ and $|\beta_n^{\#}|= o(\ln n)$.
\end{itemize}

We need to prove the above results  (ii.1)--(ii.5). Afterwards, Property (ii) of Lemma \ref{lem_Gq_cplinga} will  follow. Due to space limitation, we will detail only the proof of (ii.1), while (ii.2)--(ii.5) can be established in a way similar to those of (i.2)--(i.5).

\begin{algorithm}\setstretch{1.39}
\caption{An algorithm to find $\protect\overrightarrow{K_n^{\#}}:= [{K}_{1,n}^{\#}, {K}_{2,n}^{\#}, \ldots, {K}_{m,n}^{\#}]$ for property (ii)  of Lemma \ref{lem_Gq_cplinga}.}
\label{alg-find-K}
\begin{algorithmic}[1]
\REQUIRE $n$, $\beta_n$, $\overrightarrow{K_n}:=[K_{1,n}, K_{2,n}, \ldots, K_{m,n}]$ \vspace{2pt}
\ENSURE $\overrightarrow{K_n^{\#}}:= [{K}_{1,n}^{\#}, {K}_{2,n}^{\#}, \ldots, {K}_{m,n}^{\#}]$ \vspace{2pt}
 \STATE \textbf{let} $\widetilde{\beta_n^{\#}}   : = \min\{\beta_n, \ln \ln n\}$;\label{alg:line:betanpound-def}
 \STATE \textbf{let} $T_n:= \argmax\left\{Y: 1 -{{{P_n-Y}\choose{Y}}\over{{P_n}\choose{Y}}} \leq \frac{\ln  n + \widetilde{\beta_n^{\#}}}{n}\right\}$;\label{alg:line:tn-def}
 \IF{$K_{1,n} \geq T_n$} \label{alg:line:ifK1n} \STATEx \COMMENT{Note that we have $K_{1,n} \leq K_{2,n}\leq \ldots \leq K_{m,n}$ from the condition (\ref{K1n-K2n-Kmn-condition}).}
\FOR{each $j \in  \{1, 2, \ldots,m\}$}\label{alg:line:ifK1n1}
\STATE \textbf{let} $K_{j,n}^{\#} := T_n$;
\ENDFOR\label{alg:line:ifK1n3}
\ELSE \label{alg:line:ifK1ne}
 \STATE \textbf{let} $\ell :=  \argmax \left\{j:~\text{$1 \leq j \leq m$ and~}\text{$ K_{j,n} \leq T_n$}\right\}$;\label{alg:line:defell}
 \FOR{each $j \in  \{1, 2, \ldots,\ell\}$} \vspace{2pt}\label{alg:line:defell1}
 \STATE \textbf{let} $K_{j,n}^{\#} := K_{j,n}$;\vspace{2pt} \label{alg:line:defell2}
\ENDFOR \label{alg:line:defell3}
 \FOR{each $j \in  \{\ell+1, \ell+2, \ldots,m\}$}  \vspace{3pt} \label{alg:line:forloop}
 \STATE \textbf{let} $Q_{j,n} \hspace{-2pt} := \hspace{-2pt} \argmin\hspace{-3pt}\left\{\hspace{-4pt}Z\hspace{-2pt}:\hspace{-10pt}\begin{array}{l} \left\{\begin{array}{l} \hspace{-7pt}\hspace{-2pt}\left\{ a_j \hspace{-2pt}\left[1\hspace{-2pt} -\hspace{-2pt}{{{P_n-K_{1,n}^{\#}}\choose{Z}}\over{{P_n}\choose{Z}}}\right] \hspace{-3pt}\right\} \\ ~\\[-5pt] \hspace{-9pt}+\hspace{-2pt} \sum_{t=1}^{j-1} \hspace{-2pt}\left\{ a_t\hspace{-2pt} \left[1\hspace{-2pt} -\hspace{-2pt}{{{P_n-K_{1,n}^{\#}}\choose{K_{t,n}^{\#}}}\over{{P_n}\choose{K_{t,n}^{\#}}}}\right] \hspace{-3pt}\right\} \\ ~\\[-5pt] \hspace{-9pt}+\hspace{-2pt} \sum_{t=j+1}^{m}\hspace{-2pt} \left\{ a_t \hspace{-2pt}\left[1 \hspace{-2pt}-\hspace{-2pt}{{{P_n-K_{1,n}^{\#}}\choose{K_{t,n}}}\over{{P_n}\choose{K_{t,n}}}}\right]  \hspace{-3pt}\right\} \end{array} \hspace{-9pt}\right\}  \\ ~\\[-5pt] \geq \mathlarger{\frac{\ln  n + \widetilde{\beta_n^{\#}}}{n}}  \end{array} \hspace{-9pt}\right\}$;\vspace{3pt}\label{alg:line:argmin}
\IF{$Q_{j,n} > T_n$} \vspace{2pt} \label{alg:line:if}
\STATE \textbf{let} $K_{j,n}^{\#} := Q_{j,n}$;\vspace{2pt} \label{alg:line:if1}
 \FOR{each $r \in  \{j+1, j+2, \ldots,m\}$}\vspace{2pt} \label{alg:line:break1}
\STATE \textbf{let} $K_{r,n}^{\#} := K_{r,n}$;\vspace{2pt} \label{alg:line:break2}
\ENDFOR
\STATE \textbf{break}; \label{alg:line:break} ~~~~~~~\COMMENT{Comment: After this break statement, the execution will jump to Line \ref{alg:line:output} to output $\overrightarrow{K_n^{\#}}$.}
\ELSE \label{alg:line:else} \STATE \textbf{let} $K_{j,n}^{\#} := T_n$;  \label{alg:line:else1}
\ENDIF
\ENDFOR\label{alg:line:endfor}
\ENDIF \vspace{3pt}
\STATE \textbf{output} $\overrightarrow{K_n^{\#}}:= [{K}_{1,n}^{\#}, {K}_{2,n}^{\#}, \ldots, {K}_{m,n}^{\#}]$;\label{alg:line:output}
\end{algorithmic}
\end{algorithm}

\textit{\textbf{Proving result  (ii.1):}}

To show result  (ii.1), we will prove
\begin{align}
 K_{j,n} & \geq K_{j,n}^{\#}, \text{ for }j=1,2,\ldots,m. \label{supergraph-K-part0}
\end{align}
In Algorithm \ref{alg-find-K}, if the ``if'' statement in Line \ref{alg:line:ifK1n} is true, we  obtain (\ref{supergraph-K-part0}) from Lines \ref{alg:line:ifK1n1}--\ref{alg:line:ifK1n3} and $K_{1,n} \leq K_{2,n}\leq \ldots \leq K_{m,n}$ of the condition (\ref{K1n-K2n-Kmn-condition}). Hence, below we only need to consider the case where  the ``else'' statement in Line \ref{alg:line:ifK1ne} is executed. To this end, (\ref{supergraph-K-part0}) will be proved once the following results hold with $\ell$ defined in Line \ref{alg:line:defell}  of Algorithm \ref{alg-find-K}:
\begin{align}
 K_{j,n} & = K_{j,n}^{\#}, \text{ for }j=1,2,\ldots,\ell; \label{supergraph-K-part1new}\\
K_{\ell+1,n} & \geq K_{\ell+1,n}^{\#},   \label{supergraph-K-part1} \\ K_{J,n} & \geq K_{J,n}^{\#}, \text{ for }J=\ell+2,\ell+3,\ldots,m. \label{supergraph-K-part2}
\end{align}
 Clearly, (\ref{supergraph-K-part1new}) holds from Lines \ref{alg:line:defell1}--\ref{alg:line:defell3} of Algorithm \ref{alg-find-K}. Below we prove (\ref{supergraph-K-part2}) first and   (\ref{supergraph-K-part1}) afterwards.




\textbf{Establishing (\ref{supergraph-K-part2}).} Given an arbitrary $J\in \{\ell+2,\ell+3,\ldots,m\}$, we explain the desired result $K_{J,n} \geq K_{J,n}^{\#}$ by discussing below different cases of Algorithm \ref{alg-find-K}.
\begin{itemize}
\item[\textbf{(A)}] Here we consider the case where the ``for'' loop in Line~\ref{alg:line:forloop} of Algorithm~\ref{alg-find-K} terminates before $j$ reaches $J-1$. For example, suppose that Line~\ref{alg:line:forloop} of Algorithm~\ref{alg-find-K} is executed for only $j=\ell + 1,\ell + 2,\ldots,h$ with some integer $h $ satisfying $\ell + 1 \leq h < J-1$. Then we know that the ``break'' statement in Line \ref{alg:line:break} of Algorithm~\ref{alg-find-K} is executed for $j$ being $h$, and further know from Lines \ref{alg:line:break1} and \ref{alg:line:break2} of Algorithm~\ref{alg-find-K} that
\begin{align}
K_{t,n}^{\#} &  = K_{t,n}, \text{ for }t=h+1,h+2,\ldots,m, \nonumber
\end{align}
which with $h < J-1$ and $J \leq m$ clearly includes
\begin{align}
K_{J,n}^{\#} &  = K_{J,n}. \label{Kval-parta}
\end{align}
\item[\textbf{(B)}]  If the ``for'' loop in Line \ref{alg:line:forloop} of Algorithm \ref{alg-find-K} is now executing for $j$ being $J-1$, we divide this case to the following two cases (B1) and (B2) according to Algorithm \ref{alg-find-K}:
\begin{itemize}
 \item[\textbf{(B1)}] If $Q_{J-1,n} > T_n$, then Line  \ref{alg:line:if} of Algorithm \ref{alg-find-K} is satisfied when $j$ equals $J-1$. Thus, we obtain from Lines \ref{alg:line:if}--\ref{alg:line:break} of Algorithm~\ref{alg-find-K} that
\begin{align}
K_{t,n}^{\#} &  = K_{t,n}, \text{ for }t=J,J+1,\ldots,m,
\end{align}
which clearly includes
\begin{align}
K_{J,n}^{\#} &  = K_{J,n}.  \label{Kval-partb1-1}
\end{align}
 \item[\textbf{(B2)}] If $Q_{J-1,n} \leq T_n$,  then Line \ref{alg:line:else} of Algorithm \ref{alg-find-K} is satisfied when $j$ equals $J-1$. From Line \ref{alg:line:else1} of Algorithm \ref{alg-find-K} for $j$ being $J-1$, it holds that
\begin{align}
K_{J-1,n}^{\#} = T_{n}.  \label{Kval-partb2-1}
\end{align}
We now use the assumed condition $Q_{J-1,n} \leq T_n$ in case (B2) here. From $Q_{J-1,n} \leq T_n$ and the definition of $Q_{J-1,n}$ in Line \ref{alg:line:argmin} of Algorithm \ref{alg-find-K} when $j$ is set as $J-1$, we obtain that the expression inside ``$\argmin$'' in Line \ref{alg:line:argmin} of Algorithm \ref{alg-find-K} with $j$ set as $J-1$ and with $Z$ set as $T_n$ is   satisfied; i.e.,
\begin{align}
\left\{\begin{array}{l} \left\{ a_{J-1} \left[1 -{{{P_n-K_{1,n}^{\#}}\choose{T_n}}\over{{P_n}\choose{T_n}}}\right] \right\} \\ ~\\[-5pt] + \sum_{t=1}^{J-2} \left\{ a_t \left[1 -{{{P_n-K_{1,n}^{\#}}\choose{K_{t,n}^{\#}}}\over{{P_n}\choose{K_{t,n}^{\#}}}}\right] \right\} \\ ~\\[-5pt] + \sum_{t=J}^{m} \left\{ a_t \left[1 -{{{P_n-K_{1,n}^{\#}}\choose{K_{t,n}}}\over{{P_n}\choose{K_{t,n}}}}\right]  \right\} \end{array} \right\}   \geq \frac{\ln  n + \widetilde{\beta_n^{\#}}}{n}, \label{Kval-partb2-2}
\end{align}
where $\widetilde{\beta_n^{\#}}$ is defined in Line \ref{alg:line:betanpound-def}  of Algorithm \ref{alg-find-K}.

From (\ref{Kval-partb2-1}), the left hand side of (\ref{Kval-partb2-2}) can be written as
\begin{align}
\left\{\begin{array}{l} \left\{ a_J \left[1 -{{{P_n-K_{1,n}^{\#}}\choose{K_{J,n}}}\over{{P_n}\choose{K_{J,n}}}}\right] \right\} \\ + \sum_{t=1}^{J-1} \left\{ a_t \left[1 -{{{P_n-K_{1,n}^{\#}}\choose{K_{t,n}^{\#}}}\over{{P_n}\choose{K_{t,n}^{\#}}}}\right] \right\} \\ + \sum_{t=J+1}^{m} \left\{ a_t \left[1 -{{{P_n-K_{1,n}^{\#}}\choose{K_{t,n}}}\over{{P_n}\choose{K_{t,n}}}}\right]  \right\} \end{array} \right\}  . \label{Kval-partb2-3}
\end{align}
In case (B2) here, we have already explained that when $j$ equals $J-1$, Line \ref{alg:line:else1} of Algorithm \ref{alg-find-K} is executed. Then for $j$ being $J$, Line \ref{alg:line:forloop} of Algorithm \ref{alg-find-K} is also executed. Afterwards, for $j$ being $J$, Line \ref{alg:line:argmin} of Algorithm \ref{alg-find-K} is executed, so we define $Q_{J,n}$. From (\ref{Kval-partb2-2}) and the fact that the left hand side of (\ref{Kval-partb2-2}) equals (\ref{Kval-partb2-3}), it follows that
\begin{align}
\text{(\ref{Kval-partb2-3})} &  \geq \frac{\ln  n + \widetilde{\beta_n^{\#}}}{n}.   \label{Kval-partb2-4}
\end{align}
From (\ref{Kval-partb2-4}) and the expression in (\ref{Kval-partb2-3}), the expression inside ``$\argmin$'' in Line \ref{alg:line:argmin} of Algorithm \ref{alg-find-K} with $j$ set as $J$ and with $Z$ set as $K_{J,n}$ is   satisfied. This means
\begin{align}
K_{J,n} &  \geq  Q_{J,n},   \label{Kval-partb2-5}
\end{align}

As explained above, for $j$ being $J$, Line \ref{alg:line:forloop} of Algorithm~\ref{alg-find-K} is executed. Then from Lines \ref{alg:line:forloop}--\ref{alg:line:endfor} for $j=J$, it holds that
\begin{align}
K_{J,n}^{\#} =  \max \{Q_{J,n}, T_n\}.  \label{Kval-partb2-7}
\end{align}

From $J > \ell$, the definition of $\ell$ in Line \ref{alg:line:defell}  of~Algorithm~\ref{alg-find-K}, and the condition $K_{1,n} \leq K_{2,n}\leq \ldots \leq K_{m,n}$ from (\ref{K1n-K2n-Kmn-condition}), it holds that
\begin{align}
K_{J,n} &  \geq  T_n,   \label{Kval-partb2-6}
\end{align}

Substituting (\ref{Kval-partb2-5}) and (\ref{Kval-partb2-6}) into (\ref{Kval-partb2-7}), we know for case (B2) here,
\begin{align}
K_{J,n} \geq K_{J,n}^{\#} .  \label{Kval-partb2-8}
\end{align}

\end{itemize}
\end{itemize}

Summarizing (\ref{Kval-partb1-1}) for case (A), (\ref{Kval-partb2-1}) for case (B1), and (\ref{Kval-partb2-8}) for case (B2), in any case, we always have
\begin{align}
K_{J,n} \geq K_{J,n}^{\#} .  \label{Kval-KJn-summarize}
\end{align}

For the above analysis, we can consider any $J$ in $\{\ell+2,\ell+3,\ldots,m\}$, so we use (\ref{Kval-KJn-summarize}) to have (\ref{supergraph-K-part2}); i.e., $K_{J,n} \geq K_{J,n}^{\#}$ for $J=\ell+2,\ell+3,\ldots,m$.

\textbf{Establishing (\ref{supergraph-K-part1}).} From Lines \ref{alg:line:forloop}--\ref{alg:line:endfor} for $j=\ell+1$, it holds that
\begin{align}
K_{\ell+1,n}^{\#} =  \max \{Q_{\ell+1,n}, T_n\}.  \label{Kval-partb2-7-newv}
\end{align}

From the definition of $\ell$ in Line \ref{alg:line:defell}  of~Algorithm~\ref{alg-find-K}, and the condition $K_{1,n} \leq K_{2,n}\leq \ldots \leq K_{m,n}$ from (\ref{K1n-K2n-Kmn-condition}), it holds that
\begin{align}
K_{\ell+1,n} &  \geq  T_n,   \label{Kval-partb2-6-newv}
\end{align}
Given (\ref{Kval-partb2-7-newv}) and (\ref{Kval-partb2-6-newv}), we will have (\ref{supergraph-K-part1}) (i.e., $K_{\ell+1,n} \geq K_{\ell+1,n}^{\#}$) once proving
\begin{align}
K_{\ell+1,n} \geq Q_{\ell+1,n} .  \label{Kval-partb2-8-newv}
\end{align}

Setting $j$ as $\ell+1$ in Line \ref{alg:line:argmin} of Algorithm \ref{alg-find-K}, we obtain the definition of $Q_{\ell+1,n}$. To prove (\ref{Kval-partb2-8-newv}), it suffices to show that the expression inside ``$\argmin$'' in Line \ref{alg:line:argmin} of Algorithm \ref{alg-find-K} with $j$ set as $\ell+1$ and with $Z$ set as $K_{\ell+1,n}$ is satisfied; i.e.,
\begin{align}
\left\{\begin{array}{l} \left\{ a_{\ell+1} \left[1 -{{{P_n-K_{1,n}^{\#}}\choose{K_{\ell+1,n}}}\over{{P_n}\choose{K_{\ell+1,n}}}}\right] \right\} \\ + \sum_{t=1}^{\ell} \left\{ a_t \left[1 -{{{P_n-K_{1,n}^{\#}}\choose{K_{t,n}^{\#}}}\over{{P_n}\choose{K_{t,n}^{\#}}}}\right] \right\} \\ + \sum_{t=\ell+2}^{m} \left\{ a_t \left[1 -{{{P_n-K_{1,n}^{\#}}\choose{K_{t,n}}}\over{{P_n}\choose{K_{t,n}}}}\right]  \right\} \end{array} \right\} \geq \frac{\ln  n + \widetilde{\beta_n^{\#}}}{n} .  \label{Kval-partb2-8-newvsb}
\end{align}
Applying Line \ref{alg:line:defell2} of Algorithm \ref{alg-find-K} (i.e., $K_{j,n}^{\#} := K_{j,n}$ for $j=1, 2, \ldots,\ell$ to (\ref{Kval-partb2-8-newvsb}), we know the left hand side of (\ref{Kval-partb2-8-newvsb}) equals $\sum_{t=1}^{m} \left\{ a_t \left[1 -{{{P_n-K_{1,n}}\choose{K_{t,n}}}\over{{P_n}\choose{K_{t,n}}}}\right]  \right\}$ and hence equals $b_{1}(\overrightarrow{a}, \overrightarrow{K_n},P_n)$ from (\ref{psq1conta7tttaareae-restatedeqn}). From the condition  (\ref{lem-coupling-condition-edge-prob-beta-n}) \vspace{2pt} (i.e., $b_1(\overrightarrow{a}, \overrightarrow{K_n},P_n)=\frac{ \ln  n + \beta_n}{n}$), \vspace{2pt} it further follows that  the left hand side of (\ref{Kval-partb2-8-newvsb}) equals $\frac{ \ln  n + \beta_n}{n}$. \vspace{2pt} Then we clearly establish (\ref{Kval-partb2-8-newvsb}) from $\beta_n \geq \widetilde{\beta_n^{\#}}$, which holds from   the definition of  $\widetilde{\beta_n^{\#}}$ in Line \ref{alg:line:betanpound-def}  of Algorithm \ref{alg-find-K} (i.e., $\widetilde{\beta_n^{\#}}    = \min\{\beta_n, \ln \ln n\}$).

As explained, substituting (\ref{Kval-partb2-6-newv}) and (\ref{Kval-partb2-8-newv}) into (\ref{Kval-partb2-7-newv}), we establish the desired result (\ref{supergraph-K-part1}) (i.e., $K_{\ell+1,n} \geq K_{\ell+1,n}^{\#}$).

Finally, combining (\ref{supergraph-K-part2})   (\ref{supergraph-K-part1}) and (\ref{supergraph-K-part1new}) which we have established, we have $K_{j,n} \geq K_{j,n}^{\#}$ for $j=1,2,\ldots,m$. \vspace{2pt} Then $\mathbb{G}(n,\overrightarrow{a}, \overrightarrow{K_n},P_n)$ is a spanning supergraph of $\mathbb{G}(n,\overrightarrow{a}, \overrightarrow{K_n^{\#}},P_n)$.
Hence, result  (ii.1) is proved.

\section{Proving Theorem \ref{thm:OneLaw+NodeIsolation} for $\mathbb{G}(n,\protect\overrightarrow{a},\protect\overrightarrow{K_n},P_n)$\\under $|{\beta_n} |=  o ( \ln n)$} \label{sec-prf-thm-under-cpl}

From Section \ref{sec-confine-beta-n-thm:OneLaw+NodeIsolation-old}, we can introduce $|{\beta_n} |=  o ( \ln n)$ for proving Theorem \ref{thm:OneLaw+NodeIsolation}.
For convenience, we let {a condition set $\mathbb{C}$} denote the conditions of Theorem \ref{thm:OneLaw+NodeIsolation} with $|{\beta_n} |=  o ( \ln n)$; i.e.,
 \begin{align}
\mathbb{C}~: = & ~\big\{ \text{$ P_n = \Omega(n)$, (\ref{KPPremo}), (\ref{eq:scalinglaw_old_2}) and $|{\beta_n} |=  o ( \ln n)$} \big\} .\label{conditionsetC}
\end{align}
Our goal is to prove (\ref{thm-con-eq-0c}) and (\ref{thm-con-eq-1c}) under the condition set $\mathbb{C}$.

\subsection{Connectivity versus the absence of isolated node} \label{sec-rel-con}

In proving Theorem \ref{thm:OneLaw+NodeIsolation}, we use the relationship between connectivity and the absence of isolated node.
Clearly, if a graph is connected, then
it contains no~isolated~node \cite{ZhaoYaganGligor}. Therefore, we will obtain the zero-law (\ref{thm-con-eq-0c}) for connectivity once showing (\ref{thm-mnd-eq-0}) below, and obtain the one-law (\ref{thm-con-eq-1c}) for connectivity once showing (\ref{thm-mnd-eq-1}) and  (\ref{eq:OneLawAfterReductionsb}) below:
\begin{subnumcases}
{ \hspace{-23pt}  \lim_{n \rightarrow \infty }\hspace{-2.5pt} \mathbb{P}\hspace{-1pt}\bigg[
\hspace{-4.5pt}\begin{array}{c}
\mathbb{G}(n,\overrightarrow{a},\overrightarrow{K_n},P_n)~\mbox{has} \\
\mbox{~no~isolated~node.}
\end{array}\hspace{-5pt}
\bigg] \hspace{-3.5pt}=\hspace{-3.5pt}}  \hspace{-5pt}0,\quad\hspace{-5pt}\text{if  \hspace{-4pt}}\lim_{n \to \infty}\hspace{-1.5pt}{\beta_n}  \hspace{-1.5pt} =  \hspace{-1.5pt} - \infty, \label{thm-mnd-eq-0} \\
\hspace{-5pt}1,\quad\hspace{-5pt}\text{if  \hspace{-4pt}}\lim_{n \to \infty}\hspace{-1.5pt}{\beta_n}    \hspace{-1.5pt}=  \hspace{-1.5pt}  \infty. \label{thm-mnd-eq-1}
\end{subnumcases}
and
\begin{align}
\lim_{n \to \infty}\mathbb{P} \bigg[\hspace{-3pt}\begin{array}{c}
\mathbb{G}(n,\overrightarrow{a},\overrightarrow{K_n},P_n)\text{
has no isolated node}, \\\text{but is not connected.}
\end{array}
\hspace{-3pt}\bigg] = 0.  \label{eq:OneLawAfterReductionsb}
 \end{align}

We formally present the above result as two lemmas below.

\begin{lem} \label{lem_Gqsbsd}

{
For a graph $\mathbb{G}(n,\overrightarrow{a}, \overrightarrow{K_n},P_n)$ under the condition set $\mathbb{C}$ of (\ref{conditionsetC}), we have (\ref{thm-mnd-eq-0}) and (\ref{thm-mnd-eq-1}).
 }
\end{lem}

\begin{lem} \label{lem_Gq_no_isolated_but_not_conn}

{
For a graph $\mathbb{G}(n,\overrightarrow{a}, \overrightarrow{K_n},P_n)$ under the condition set $\mathbb{C}$ of (\ref{conditionsetC}), we have (\ref{eq:OneLawAfterReductionsb}).

 }
\end{lem}
Lemma \ref{lem_Gqsbsd} presents a zero--one law for the absence of isolated node via  (\ref{thm-mnd-eq-0}) and (\ref{thm-mnd-eq-1}).
In the rest of this section, we discuss the proofs of Lemmas \ref{lem_Gqsbsd} and \ref{lem_Gq_no_isolated_but_not_conn}, respectively. {We will often write $\mathbb{G}(n,\overrightarrow{a}, \overrightarrow{K_n},P_n)$ as $\mathbb{G}$} for   brevity.

 \subsection{Proof of Lemma \ref{lem_Gqsbsd}} \label{sec-lem_Gqsbsd}

 To prove Lemma \ref{lem_Gqsbsd} on the existence/absence of isolated node, we use the method of moments \cite{ZhaoYaganGligor} to evaluate the number of of isolated nodes. The proof idea is similar to those by
Ya{\u{g}}an
\cite{yagan-heter-IT} and Zhao \textit{et al.} \cite{ZhaoYaganGligor}.

First, we will prove (\ref{thm-mnd-eq-0}) by showing that  ${I_n}$,  denoting the number of  nodes that belong to group $\mathcal{A}_1$ and are isolated in $\mathbb{G}$ (i.e., $\mathbb{G}(n,\overrightarrow{a},\overrightarrow{K_n},P_n)$), is positive \emph{almost surely}, where an event (indexed by $n$)
 {occurs \emph{almost surely} if its probability converges to $1$ as
$n\to \infty$}. Formally, $\lim_{n \to \infty} \bP{  {I_n}  >0} = 1$ or equivalently $\lim_{n \to \infty} \bP{  {I_n}  = 0 } = 0$. The inequality $\bP{  {I_n}  = 0 } \leq 1 - { \bE{ {I_n} }^2}\big/{
\bE{ {I_n}  ^2} }$ holds from the method of second moment \cite{ZhaoYaganGligor}, so proving (\ref{thm-mnd-eq-0}) reduces to showing $\lim_{n \to \infty} { \bE{ {I_n} }^2}\big/{
\bE{ {I_n}  ^2} } = 1$. With indicator variables $\psi_{n,i}$ for $i=1, \ldots , n$ denoting $\1{ {\rm Node~}v_i~\text{belongs to group $\mathcal{A}_1$ and is~isolated~in~} \mathbb{G} . },$ we have ${I_n}  = \sum_{i=1}^n \psi_{n,i}$. Noting that the random variables $\psi_{n,1}, \ldots , \psi_{n,n} $ are
exchangeable due to symmetry, we find
$\bE{ {I_n} } = n \bE{ \psi_{n,1}  }$
and
$\bE{ {I_n} ^2 } = n \bE{{\psi_{n,1}}^2   }
 +  n(n-1) \bE{ \psi_{n,1}   \psi_{n,2}  } = n \bE{ \psi_{n,1}  }
 +  n(n-1) \bE{ \psi_{n,1}   \psi_{n,2}  },$
where the last step uses $\bE{{\psi_{n,1}}^2   } = \bE{ \psi_{n,1}  }$
as $\psi_{n,1}$ is a binary random variable. It then follows that $
\frac{ \bE{ {I_n} ^2 }}{ \bE{ {I_n}  }^2 } = \frac{
1}{  n\bE{ \psi_{n,1}  } }
  + \frac{n-1}{n} \cdot \frac{\bE{ \psi_{n,1}
\psi_{n,2}  }}
     {\left (  \bE{ \psi_{n,1}  } \right )^2 }.$ Given this and the standard inequality $ \bE{ {I_n} ^2 } \geq \bE{ {I_n}  }^2$, we will obtain $\lim_{n \to \infty} { \bE{ {I_n} }^2}\big/{
\bE{ {I_n}  ^2} } = 1$ and thus the desired result (\ref{thm-mnd-eq-0}) once proving  \begin{align}
\lim_{n \to \infty} \big(n \bE{ \psi_{n,1}  }\hspace{-.5pt}\big) &= \infty\text{ if $\lim_{n \to \infty} {\beta_n}   =   - \infty$},\text{ and}
\label{eq:OneLaw+NodeIsolation+convergence2} \\
\hspace{-5pt}  {\bE{ \psi_{n,1}
\psi_{n,2}  }}\big/{\left (  \bE{ \psi_{n,1}  }\hspace{-.5pt} \right )^2 }
 &\leq 1 + o(1)\text{ if $\lim_{n \to \infty} {\beta_n}   =   - \infty$}. \label{eq:ZeroLaw+NodeIsolation+convergence}
\end{align}

Second, we will prove (\ref{thm-mnd-eq-1}) by showing that  ${J_n}$,  denoting the number of isolated nodes in $\mathbb{G}$, is zero almost surely; i.e., $\lim_{n \to \infty} \bP{  {J_n}  = 0} = 1$. The inequality $1 - \bE{ {J_n}  } \leq \bP{  {J_n}  = 0 }$ holds from the method of first moment \cite{ZhaoYaganGligor}, so proving (\ref{thm-mnd-eq-0}) reduces to showing $\lim_{n \to \infty} \bE{ {J_n}  }  = 0$. With indicator variables $\phi_{n,i}$ for $i=1, \ldots , n$ denoting $\1{ {\rm Node~}v_i~\text{is~isolated~in~} \mathbb{G} . },$ we have ${J_n}  = \sum_{i=1}^n \phi_{n,i}$. Noting that the random variables $\phi_{n,1}, \ldots , \phi_{n,n} $ are
exchangeable due to symmetry, we find
$\bE{ {J_n} } = n \bE{ \phi_{n,1}  }$. Given the above, we will obtain the desired result (\ref{thm-mnd-eq-1}) once proving  \begin{equation}
\lim_{n \to \infty} \big(n\bE{ \phi_{n,1}  }\hspace{-.5pt}\big)= 0\text{ if $\lim_{n \to \infty} {\beta_n}   = \infty$}.
\label{eq:OneLaw+NodeIsolation+convergence-one}
\end{equation}

As explained above, proving Lemma \ref{lem_Gqsbsd} reduces to showing (\ref{eq:OneLaw+NodeIsolation+convergence2}) (\ref{eq:ZeroLaw+NodeIsolation+convergence}) and (\ref{eq:OneLaw+NodeIsolation+convergence-one}). Their proofs have been discussed in the conference version \cite{Zhao-GlobalSIP} and are similar to those by
Ya{\u{g}}an
\cite{yagan-heter-IT} and Zhao \textit{et al.} \cite{ZhaoYaganGligor} (still we tackle a more general set of parameter conditions and a more fine-grained scaling than \cite{yagan-heter-IT}). Due to space limitation, the details are provided in \cite{fullpaper}.

 \subsection{Proof of Lemma \ref{lem_Gq_no_isolated_but_not_conn}} \label{sec-lem_Gq_no_isolated_but_not_conn}

 The goal is to show a negligible (i.e., $o(1)$) probability for  $F_n$ denoting the event that graph $\mathbb{G}$
(i.e., $\mathbb{G}(n,\overrightarrow{a},\overrightarrow{K_n},P_n)$) has no isolated node, but is not connected. The idea is to analyze the topological feature of $\mathbb{G}$ under $F_n$ \cite{yagan-heter-IT,ZhaoYaganGligor}: if $F_n$ occurs, there exists a subset $T$ of nodes with $2 \leq |T| \leq \lfloor \frac{n}{2} \rfloor $ such that $\mathcal{C}_{r,n} $ and $\mathcal{D}_{r,n}$ both happen, where
\begin{align}
\mathcal{C}_{T,n}: &  \begin{array}{l} \text{The event that
$\mathbb{G}(T)$ (i.e., the subgraph of $\mathbb{G}$ with}\\\text{the vertex set restricted to $T$) is connected},  \end{array} \label{CTndefn}
\end{align}
\begin{align}
\mathcal{D}_{T,n}:  &\begin{array}{l} \text{The event that there is no edge between any node}\\\text{in $T$ and any node in $\{v_1,v_2\ldots,v_n\}\setminus T$} . \end{array}\label{DTndefn}
\end{align}
 To get $\bP{F_n}=o(1)$,
by a union bound, it suffices to show
 \begin{align}\textstyle{\sum_{\begin{subarray}{l}T\subseteq \{v_1,v_2\ldots,v_n\}: \\ 2 \leq |T| \leq \lfloor \frac{n}{2} \rfloor \end{subarray}}\bP{\mathcal{C}_{T,n} \cap \mathcal{D}_{T,n}}=o(1).}\label{overviewprf1}
\end{align} We find that given $n$, for any $T$ with fixed $|T|=r$, $\mathcal{C}_{T,n}$ (resp., $\mathcal{D}_{T,n}$) is the same stochastically with $\mathcal{C}_{\{ v_1, \ldots , v_r \},n}$ (resp., $\mathcal{D}_{\{ v_1, \ldots , v_r \},n}$) (denoted by $\mathcal{C}_{r,n}$ and $\mathcal{D}_{r,n}$ with a little abuse of notation), so by a union bound, it suffices to establish
 \begin{align}\textstyle{\sum_{r=2}^{\lfloor n/2\rfloor} \binom{n}{r} \bP{\mathcal{C}_{r,n} \cap \mathcal{D}_{r,n}}=o(1),}\label{overviewprf2}
\end{align}   (this is not we will prove precisely, but it gives the intuition).

The rest of the proof is similar to those by
Ya{\u{g}}an
\cite{yagan-heter-IT} and Zhao \textit{et al.} \cite{ZhaoYaganGligor} (still our proof addresses a more general set of parameter conditions and a more fine-grained scaling than \cite{yagan-heter-IT}). Due to space limitation, we present the details in \cite{fullpaper}.



 \section{Conclusion} \label{sec:Conclusion}

We derive a sharp zero--one law for connectivity in a heterogeneous secure sensor network. The paper improves the seminal work \cite{yagan-heter-IT} of Ya{\u{g}}an since our zero--one law applies to a more general set of parameters and is more fine-grained. Our work provides useful guidelines for designing  secure sensor networks under a heterogeneous key predistribution scheme.

\normalsize
 
\end{document}